\documentclass[12pt,a4paper]{article}
\usepackage{jheppub}
\usepackage{multirow}
\usepackage{chemarrow}
\usepackage{amsmath,amssymb,amsfonts,mathtools}
\usepackage{float}
\usepackage{subfig}
\usepackage{scalefnt}
\usepackage{wrapfig}
\usepackage{fancybox}
\usepackage{ifsym}
\usepackage{relsize}
\newcommand{\bea}{\begin{eqnarray}\displaystyle}
\newcommand{\eea}{\end{eqnarray}}

\newlength{\arrow}
\newcommand\half{\frac{1}{2}}

{\setlength{\fboxsep}{15pt}
\setlength{\mylength}{\linewidth}%
\addtolength{\mylength}{-2\fboxsep}%
\addtolength{\mylength}{-2\fboxrule}%
\Sbox
\minipage{\mylength}%
\setlength{\abovedisplayskip}{0pt}%
\setlength{\belowdisplayskip}{0pt}%
\equation}%
{\endequation\endminipage\endSbox
\[\fbox{\TheSbox}\]}

\begin{document}
\title{On orbifolds of M-strings}
\author[\ast\ddagger]{Babak Haghighat,}
\author[\dag]{Can Koz\c{c}az,}
\author[\ast]{Guglielmo Lockhart,}
\author[\ast]{Cumrun Vafa}
\affiliation[\ast]{Jefferson Physical Laboratory, Harvard University, Cambridge, MA 02138, USA}
\affiliation[\ddagger]{Department of Mathematics, Harvard University, Cambridge, MA 02138, USA}
\affiliation[\dag]{International School of Advanced Studies (SISSA),
via Bonomea 265, 34136 Trieste, Italy and INFN, Sezione di Trieste}

\abstract{
We consider M-theory in the presence of $M$ parallel M5-branes probing a transverse $A_{N-1}$
singularity.  This leads to a superconformal theory with $(1,0)$ supersymmetry in six dimensions.
We compute the supersymmetric partition function of this theory on a two-torus, with arbitrary supersymmetry preserving twists, using the topological vertex formalism.   Alternatively, we show that this can also be obtained by computing the elliptic genus of an orbifold of recently studied M-strings.  The resulting 2d theory is  a (4,0) supersymmetric quiver gauge theory whose Higgs branch corresponds to strings propagating on the moduli space of $SU(N)^{M-1} $ instantons
on ${\mathbb R}^4$ where the right-moving fermions are coupled to a particular bundle.}
\maketitle

\section{Introduction}

The study of six-dimensional superconformal theories is still in its infancy.  We know of the existence
of $(2,0)$ and $(1,0)$ superconformal theories, and we believe we have a full list of the
$(2,0)$ theories, classified by ADE-type.  The classification of $(1,0)$ theories remains more mysterious.
There are some examples known \cite{Seiberg:1996vs,Ganor:1996mu,Blum:1997mm}.  In particular the paper~\cite{Blum:1997mm} considers 5-branes in various theories probing transverse ADE-type singularities and argues that this leads to $(1,0)$ superconformal theories in six dimensions.   In this paper
we study the partition function of $(1,0)$ superconformal theories corresponding to $M$ M5 branes
probing transverse ADE singularities.  More specifically, in this paper we focus on the case of 
transverse $A_{N-1}$ singularities.  This system is known to be dual to type IIB strings with $N$ D5-branes
probing transverse $A_{M-1}$ singularity.

This theory has a deformation away from the conformal fixed point where the M5-branes are separated
in the extra transverse direction.  The separation between adjacent branes correspond to vevs of scalars in the $M-1$
$(1,0)$ tensor multiplets.  The presence of tensor multiplets suggests the existence
of strings charged under the 2-form fields.   The main question of interest in this paper is to investigate
to what extent these strings capture the supersymmetric partition function of this theory, along the lines
recently investigated in  \cite{Haghighat:2013gba} for the $(2,0)$ theory.  We shall see that indeed they capture the full supersymmetric
partition function of the theory on $T^2$ with arbitrary twists preserving supersymmetry.  In fact
these strings support a $(4,0)$ supersymmetric quiver gauge theory, whose
elliptic genus captures the partition function of the bulk theory.
This result also applies to the case studied in \cite{Haghighat:2013gba} as a special case
(setting N=1).

  The presence of transverse $A_{N-1}$ singularity suggests
that we have, in addition, an $SU(N)$ gauge symmetry.  This would be the case if there were no M5-branes.
However in the presence of $M$  M5-branes, the gauge symmetry turns out to enhance to
$SU(N)^{M-1}$ of an affine $A_{M-1}$ quiver  gauge theory with bifundamental matter fields with  an extra $SU(N)$ being a global symmetry \cite{Blum:1997mm}.
The easiest way to see this fact is to go to a dual type IIB description where this
corresponds to having $N$ D5-branes probing a transverse $A_{M-1}$ singularity\footnote{The absence of $U(1)^{M-1}$'s in the gauge factor is because they are anomalous and are higgsed by the hypermultiplets corresponding to the $A_{M-1}$ hyperK\"ahler moduli.}.
 In the M-theory setup we have M2-branes stretched between parallel M5-branes which
 lead to M-strings (see \cite{Haghighat:2013gba} for detailed discussion).  From the viewpoint of M-strings placing the M5-branes in the presence of $A_{N-1}$ singularity can be interpreted
 as follows:  It corresponds
 to placing $N$ copies of M-strings and modding out by a $\mathbb{Z}_{N}$ action which permutes them
 but at the same time acts by a $\mathbb{Z}_N$ subgroup of the global $SO(4)_\perp$ symmetry
 which the strings enjoy.  The main goal of this paper is to study how this orbifold
 action is perceived by the M-strings.
 
 To this end we study further compactification of this theory on $S^1$ and $S^1\times S^1$.
 As we go down to 5-dimensions on an $S^1$, we can turn on $(N-1)(M-1)$ Wilson lines of
 $SU(N)^{M-1}$ and
 the $N-1$ fugacities from the global $SU(N)$ symmetry, giving a total of $(N-1)M$ parameters.  In addition the theory depends on the 6d vev
 of the $M-1$ tensor multiplets as well as the radius of the circle.  Moreover
 as we go around the circle we can act by a supersymmetry-preserving
 transverse
 rotation, leading to a mass parameter for the
 bifundamental fields. Altogether, this gives $NM+1$ parameters.  In other words, we end up with an ${\cal N}=1$ 
 supersymmetric gauge theory 
 in 5d, which depends on these parameters (which partly specify the Coulomb branch
 of the theory and partly the coupling parameters).  One can then compute the supersymmetric
 partition function of these theories, either using the topological vertex formalism or the instanton
 calculus (which corresponds to the twisted partition function on a further compactification
 on $S^1$).  As is well known \cite{Gopakumar:1998ii,Gopakumar:1998jq,Hollowood:2003cv} these capture BPS degeneracies of the theory, which can be interpreted as arising from the M-strings \cite{Haghighat:2013gba}.  In particular the computation of the partition function of the resulting 5d theory is equivalent to computation of the elliptic genus of the corresponding strings, which in turn can be interpreted as the 
 elliptic genus of the $\mathbb{Z}_N$ orbifold of $N$ M-strings, which ends up being given by a $(4,0)$ supersymmetric quiver gauge theory.
  
 The organization of this paper is as follows:  In Section 2, we introduce the basic setup of M5-branes
  and M-strings
 in the presence of $A_{N-1}$ singularities, and we discuss the further compactifications
 on $S^1$ and $S^1\times S^1$ and the interpretation of this system in various duality frames.
 We also present the quiver description of such M-strings using the type IIB setup.  In Section 3
 we show how refined topological strings can be used to compute the partition function of this theory.
 We find that the basic building block of this computation can be interpreted in terms of the amplitudes
 of a collection of M2-branes which end on two sides of M5-branes, in the presence
 of $A_{N-1}$ singularity.  In other words the presence of M5-branes can be
 viewed as a domain wall which acts as an operator on the states of the M2-branes on the
 left, to give the states of M2-branes on the right.  We also discuss the modular properties
 of the partition functions of the theory with respect to the elliptic modulus of the $T^2$ compactification.
 We also show how these results can also be directly obtained from the quiver (4,0) gauge theory.
 In Section 4 we end with some concluding remarks.
 
We understand that related results have been obtained independently in \cite{iqbal}. We thank the authors for communicating this to us.
 
\section{Geometry of $\mathbf{M_A}$-strings}

$\mathrm{M_A}$-strings arise from M2-branes ending on parallel M5-branes in the presence of $A_{N-1}$ singularities. In this section we clarify the details of the geometry behind this construction and discuss twisted compactifications on $S^1$ and $S^1 \times S^1$. We then proceed to describe various dual descriptions of this system. In particular, by compactifying the M5-branes on $S^1$ with twisted boundary conditions we end up with a theory in five dimensions with the same degrees of freedom as a quiver version of $\mathcal{N}=2^*$. This theory has further realizations in terms of a $(p,q)$-fivebrane web in type IIB string theory as well as compactifications of M-theory on certain non-compact Calabi-Yau manifolds.

In Section \ref{sec:basics} we present the basic geometry and setup of our notation, including
how the M-strings fit in this picture, and what their global symmetries are.  In Section \ref{sec:s1} we discuss compactification on a circle and twisting around the circle to introduce a mass parameter.  In Section \ref{sec:duality} we discuss the various duality frames:  In Section \ref{sec:5d} we provide a dual type IIA description involving D4-branes probing $A_{N-1}$ singularities and its T-dual IIB description
involving a web of $(p,q)$-fivebranes as well as the corresponding toric description characterizing
M-theory on local Calabi-Yau three-folds.  In Section \ref{sec:6d} we provide yet another dual
type IIB description involving D5-branes probing $A_{M-1}$ singularities.  In Section \ref{sec:s1s1} we consider further compactification on $S^1$ which allows us to introduce the Omega background.  We also recall the refined topological string description of the partition function and its connection
with BPS degeneracies.  In Section \ref{sec:quiver} we provide the quiver description for the orbifold of M-strings  (i.e. $\mathrm{M_A}$-strings) giving a $(4,0)$ supersymmetric system which is deduced from the type IIB dual description discussed in Section \ref{sec:6d}.
In that section we point out the interpretation of the quiver theory as a gauge system whose
Higgs branch describes the moduli space of instantons on $SU(N)^{M-1}$, where the fermions are coupled to suitable bundles.

\subsection{Basics of the setup}
\label{sec:basics}
Consider $M$ parallel and coincident M5-branes in the presence of an $A_{N-1}$ singularity in the transverse directions. That is, the M5-branes fill a subspace $\mathbb{R}^6$ of $\mathbb{R}^{1,10}$, whereas the transverse space is of the form
\begin{equation}
	\mathbb{R} \times A_{N-1}, \textrm{~with~} A_{N-1} \equiv \mathbb{C}^2/\Gamma_{N}, \quad \Gamma_N = \left\{\left(\begin{array}{cc}e^{\frac{2\pi i}{N}} & 0 \\ 0 & e^{-\frac{2\pi i}{N}}\end{array}\right) | i = 1, \cdots, N-1\right\}.
\end{equation}
The space on which M-theory is compactified is then $\mathbb{R}^6_{||}\times \mathbb{R} \times (A_{N-1})_{\perp}$, where the subscripts are used to distinguish directions parallel or transverse to the worldvolume of the M5-branes.
The resulting theory living on the M5-branes then has $(1,0)$ supersymmetry. The massless representations of this supersymmetry are then labeled by their $Spin(4) \sim SU(2)^\parallel_L \times SU(2)^\parallel_R$ representations. Scalars arise from hypermultiplets as well as from the tensor multiplets. 

We choose coordinates $X^I,~ I=0,1,2,\cdots,10$, and parametrize the worldvolume of the M5-branes by $X^0,X^1,X^2,X^3,X^4,X^5$. We take the transverse $\mathbb{R}^4$, which we mod out by the orbifold group $\Gamma_N$, to be parametrized by $X^7,X^8,X^9,X^{10}$ which we also sometimes denote by $\mathbb{R}^4_{\perp}$. Next, we separate the M5-branes along the $X^6$ directions and denote their position in the $X^6$ direction by $a_i, i =1,2,\cdots,M$. Thus, before orbifolding, rotations of $\mathbb{R}^4_{\perp}$ will lead to a $Spin_R(4)\sim SU(2)_L^\perp\times SU(2)_R^\perp\,$ R-symmetry on the M5-brane worldvolume theory. Following \cite{Haghighat:2013gba}, one can introduce M2-branes ending on M5-branes with boundary coupling to the anti-symmetric 2-form field, whose worldvolume is along the $X^0,X^1,X^6$ directions. Altogether we have the following setup:
\begin{equation}
	\begin{array}{c|ccccccccccc}
		~ & X^0 & X^1 & X^2 & X^3 & X^4 & X^5 & X^6 & X^7 & X^8 & X^9 & X^{10} \\
		\hline 
		\mathbb{C}^2/\Gamma_N & - & - & - & - & - & - & - & \times & \times & \times & \times \\
		\textrm{M5} & \times & \times & \times & \times & \times & \times & \{a_i\} & - & - & - & - \\
		\textrm{M2} & \times & \times & -         & -         & -         & -         & \times  & - & - & - & -
	\end{array}
\end{equation} 
The boundaty of an M2-brane inside an M5-brane is spanned by $(X^0,X^1)$ and is a string inside the M5-brane, which following the terminology of \cite{Haghighat:2013gba} we now call a $\textrm{M}_\textrm{A}$-string as there is an $A_{N-1}$ singularity transverse to the fivebrane. The presence of the string breaks the $Spin(1,5)$ Lorentz symmetry of the M5-brane to $Spin(1,1)\times Spin(4)$,  $Spin(1,1)$ being the Lorentz group on the string. As shown in \cite{Haghighat:2013gba} the chiralities of the preserved supersymmetries on the M-string under $Spin(1,1)$, $Spin_R(4)$ and $Spin(4) \subset Spin(1,5)$  are equal. Thus before the $\Gamma_N$ orbifold action the preserved supersymmetries organize themselves into four left-moving and four right-moving supercharges whose eigenvalues under 
\begin{equation}
	Spin(4) \sim SU(2)^{||}_{L} \times SU(2)^{||}_{R} , \quad Spin_R(4) \sim SU(2)^{\perp}_{L} \times SU(2)^{\perp}_{R},
\end{equation}
and are given in table \ref{tab:supercharges}.
\begin{table}[h!] 
\begin{center}
	$\begin{array}{cccc|cccc}
		\multicolumn{4}{c|}{L} & \multicolumn{4}{|c}{R} \\
		\hline
		 J^{||,L}_{3} &  J_3^{||,R} &  J_3^{\perp,L} & J_3^{\perp,R} & J_3^{||,L} &  J_3^{||,R} &  J_3^{\perp,L} & J_3^{\perp,R}\\
		 \hline
		 + & + & - & - & + & - & - & +\\
		 - & - & + & + & - & + & + & - \\
		 - & - & - & - & + & - & + & - \\
		 + & + & + & + & - & + & - & + \\
	\end{array}$
\end{center}
	\caption{Preserved supersymmetries on the string before $\mathbb{Z}_N$ orbifold action. The table shows the Cartan eigenvalues of $SO(8)$ where it is implicit that all signs are multiplied by $\half$. The two columns of the table correspond to the left-moving and right-moving supercharges on the worldsheet of the M-string.}
	\label{tab:supercharges}
\end{table}

Note that these supercharges form a positive chirality spinor of $Spin(8)$, namely $\mathbf{8}_s$. It is now easy to include the action of the orbifold group. For this we note that supercharges transform under the action of the orbifold group as
\begin{equation} 
	Q_{\mathbf{s}} \mapsto \exp(2\pi i \mathbf{s} \cdot \vec{\zeta}) Q_{\mathbf{s}},
\end{equation}
where $\vec{\zeta}=(0,0,\zeta_1,\zeta_2)$ parametrizes the orbifold action which in our case is given by
\begin{equation}\label{eq:orbaction}
	(w_1,w_2) \in \mathbb{C}^2 \simeq \mathbb{R}^4_{\perp} \Rightarrow \Gamma_N: (w_1,w_2) \mapsto (e^{2 \pi i \zeta_1} w_1,e^{2\pi i \zeta_2} w_2),
\end{equation}
with $\zeta_1=\frac{1}{N}$ and $\zeta_2 = -\frac{1}{N}$. Therefore, we see that only the left-moving supercharges survive as they are the only ones which are invariant under the action (\ref{eq:orbaction}). This shows that the worldvolume supersymmetry is reduced from $ (4,4) $ to $(4,0)$ by the orbifolding.

\subsection{Compactification on $S^1$ and mass rotation}
\label{sec:s1}
Next, we consider compactifying $X^1$ to a circle of radius $R_1$. Recall that the transverse $\mathbb{R}^4$ is parametrized by $X^7,X^8,X^9,X^{10}$ and is modded out by the orbifold group $\Gamma_N$ to give an $A_{N-1}$ singularity. Resolving this singularity gives rise to an ALE space with metric
\begin{eqnarray} \label{eq:ALE}
	ds^2 & = & V^{-1} (dt+\vec{A}\cdot d \vec{x})^2 + V d\vec{x}^2 \nonumber \\
	V       & = & \sum_{i=1}^N \frac{1}{|\vec{x} - \vec{x}_i|} \nonumber \\
	- \vec{\nabla} V & = & \vec{\nabla} \times \vec{A}~.
\end{eqnarray}
The second homology of this space is generated by two-cycles $C_i$, $i=1,\cdots,N-1$ whose intersection numbers produce the Cartan matrix of $A_{N-1}$. This space can be equivalently viewed as a specific limit of the multi-centred Taub-Nut space $\textrm{TN}_N$ defined by the same equations as above with the modification that $V$ gets replaced by 
\begin{equation}
	V = \sum_{i=1}^N \frac{1}{|\vec{x} - \vec{x}_i|} + \frac{1}{\lambda^2}.
\end{equation}
The underlying geometry is then a circle fibration over $\mathbb{R}^3$ such that the circle shrinks to zero size at the points $\vec{x}_i \in \mathbb{R}^3$ and approaches an asymptotic value at infinity, namely $\lambda$. In the limit $\lambda \rightarrow \infty$ one then regains the ALE space (\ref{eq:ALE}). However, for our purposes, when we talk about the $A_{N-1}$ singularity we will always keep the circle at infinity finite and therefore will consider $\textrm{TN}_{N}$ in this paper.

Let us next come to the isometries of the space $\textrm{TN}_N$. Generically, the isometry group is just $U(1)_f$, corresponding to rotation of the circle fiber. Furthermore, for configurations where all centers are aligned along a line there is another $U(1)$ isometry which corresponds to rotations preserving this axis, denoted by $U(1)_b$\footnote{For $N=1$ this isometry gets enhanced to $SU(2)$ and thus the full isometry group of $\textrm{TN}_1$ is $U(1)_f \times SU(2)_b$.}. The situation is analogous to the isometries of $A_{N-1}$ ALE space discussed in \cite{Gibbons:1996nt}. We want to describe both $U(1)$'s explicitly by choosing complex coordinates. To this end, we recall that the singular limit of this space corresponds locally around the origin to the algebraic equation
\begin{equation}
	X^N + Y Z =0 
\end{equation}
in $\mathbb{C}^3$. We can  parametrize solutions by $Y = w_1^N$, $Z= w_2^N$ and $X=w_1 w_2$. Note that these equations are preserved when blowing up the singularity and are therefore a valid description of $\textrm{TN}_N$ around the origin. The two isometries discussed above then have the following representations in this picture:
\begin{eqnarray}
	U(1)_f & : & (w_1,w_2) \mapsto (e^{2\pi i \alpha} w_1, e^{-2\pi i \alpha} w_2) \nonumber \\
	U(1)_b & : & (w_1,w_2) \mapsto (e^{2\pi i \alpha} w_1, e^{2\pi i \alpha} w_2).
\end{eqnarray}

\begin{figure}[here!]
  \centering
	\includegraphics[width=0.5\textwidth]{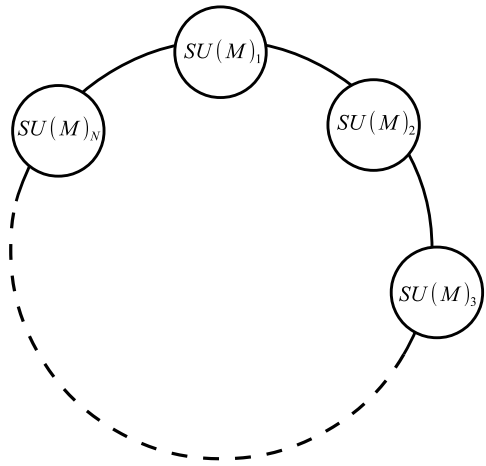}
  \caption{Compactification of the M5-brane theory on a circle in the presence of an $A_{n-1}$ singularity leads to the 5d quiver gauge theory depicted here.}
  \label{fig:5dquiver1}
\end{figure}

Having identified the isometries of the space transverse to the M5-branes we next consider compactification of the coordinate $X^1$ on a circle with radius $R_1$. We can fiber $\textrm{TN}_N$ non-trivially over this $S^1$ as follows: as we go around the circle we use the isometry $U(1)_f$ to rotate $(w_1,w_2)$,
\begin{equation}
	U(1)_m \equiv U(1)_f : (w_1,w_2) \rightarrow (e^{2\pi i m} w_1, e^{-2\pi i m} w_2).
\end{equation}
Note that the supercharges that are invariant under this rotation are precisely the left-moving supercharges that survive the orbifold action \eqref{eq:orbaction}. For $N=1$, the resulting theory in $5d$ is an $\mathcal{N}=2^*$ theory with $SU(M)$ gauge group and adjoint hypermultiplet with mass $m$ which was studied in \cite{Haghighat:2013gba}. For general $N$ the theory is an affine $\hat{A}_{N-1}$ quiver gauge theory with an $SU(M)$ gauge group at each node and with bi-fundamental matter between adjacent nodes. We depict this in Figure \ref{fig:5dquiver1}.
There are $N$ different gauge couplings, one for each node in the quiver, and their sum is related to the radius of the circle along the $X^1$ direction through
\begin{equation}
	\tau = \sum_{i=1}^N \tau_i = \sum_{i=1}^N \frac{4 \pi^2}{g_{YM,i}^2} = \frac{1}{R_1},
\end{equation}
where we take the $\tau_i$ to be the couplings of the individual nodes. Furthermore, the hypermultiplets which form the bi-fundamental matter fields will each have mass $m$. To complete the count of parameters note that there are also $N(M-1)$ Coulomb branch parameters. Together with the mass parameter and the couplings we thus see that the gauge theory depends altogether on $N(M-1) + N + 1 = NM+1$ parameters.

\subsection{Different duality frames}
\label{sec:duality}
In this section we present different realizations within type II string theory of the M-theory setup discussed above. The goal will be to derive on the one hand a type IIB $(p,q)$-brane web construction for the 5d gauge theory which will allow us to lift the brane setup to a M-theory compactification on a non-compact Calabi-Yau threefold. On the other hand we will derive another type IIB description in terms of D5-branes in the presence of $A_{M-1}$ singularity which will serve two purposes. First of all, it will give rise to a dual 6d gauge theory description of the original M-theory setup, and secondly it will allow us to give a 2d quiver gauge theory description for the $\textrm{M}_{\textrm{A}}$-strings. 

\subsubsection{Type IIB $(p,q)$-brane web and M-theory on toric Calabi-Yau}
\label{sec:5d}
Let us start with the derivation of the type IIB $(p,q)$-fivebrane web setup through a chain of dualities.
As a first step we compactify the original M-theory geometry along the $ X^1 $ circle. We obtain type IIA theory with the following brane setup:
\begin{center}
\begin{tabular}{c|cccccccccc}
& $ \mathbb{R} $ & \multicolumn{4}{c}{$\mathbb{R}^4_\parallel$}&$\mathbb{R}$ &\multicolumn{4}{c}{TN$_N$}\\
&  $ X^0 $& $ X^2 $& $ X^3 $& $ X^4 $& $ X^5 $& $ X^6 $& $ X^7 $& $ X^8 $& $ X^9 $ & $X^{10}$\\
\hline
M $D4$ & $\times$ & $\times$ & $\times$ &$\times$ &$\times$ & $\{a_i\}$ & -- & -- & --& --\\
k $F1$ & $\times$ & $-$ &$-$ &$-$ &$-$ &  $\times$ & -- & -- & -- & --\\
\end{tabular}
\end{center}
That is, we have $ k $ fundamental strings stretching between D4-branes, in a transverse Taub-NUT background of charge $ N $. We denote the separations between the D4-branes by $t_f^{i}$, while $ \tau $ is now related to the gauge coupling of the D4-brane worldvolume theory by
\[ \frac{g_{YM}^2}{4\pi^2}= \tau^{-1}.\]
The presence of transverse $A_{N-1}$-singularity leads to a $\mathbb{Z}_N$ orbifold \cite{Douglas:1996sw} and this gives rise to the five-dimensional quiver gauge theory described in the previous section. Let us next discuss the reduction of the M-theory 3-form $A^{(3)}$. Before the circle-reduction it can be given an expectation value along the three-cycles $S^1 \times C_i$ where $S^1$ is the M-theory circle. These particular expectation values will reduce in the type IIA setup to non-zero B-field flux on the $ C_i $ cycles:
\[ B = \sum_{i=1}^{N-1}\tau_i\, \cdot\, \omega_i \rightarrow \int_{C_i} B = \tau_i,\]
where we take the $\omega_i$ to be elements of $H^{1,1}(\textrm{TN}_N,\mathbb{Z})$ and Poincare dual to the $C_i$. 

Let us next assume that $ m $ is turned off\footnote{The mass parameter, which had entered as a twist along $ X^1 $ of the transverse $\textrm{TN}_N$, now has the following interpretation: upon compactifying on $ X^1 $, we get a new gauge field $ A_m $ from the metric: $ A_m = g_{1 \theta} = m\, d\theta $, where $ \theta $ parametrizes the Taub-NUT fiber. Thus we find that there is a nonzero Wilson line along the Taub-Nut fiber:
$\oint_\theta A_m = 2\pi i m$.}. Now we perform T-duality along the Taub-NUT circle.  The Taub-NUT geometry turns into a collection of type IIB NS5-branes on transverse $ S^1\times \mathbb{R}^3 $ \cite{Ooguri:1995wj}, while the D4-branes become D5-branes and the fundamental strings of type IIA turn into type IIB fundamental strings. We end up with the following picture:
\begin{center}
\begin{tabular}{c|cccccccccc}
& $ \mathbb{R} $ & \multicolumn{4}{c}{$\mathbb{R}^4_\parallel$}&$\mathbb{R}$ & $S^1$ &\multicolumn{3}{c}{$\mathbb{R}^3_\perp$}\\
& $ X^0 $& $ X^2 $& $ X^3 $& $ X^4 $& $ X^5 $& $ X^6 $& $ X^7 $& $ X^8 $& $ X^9 $& $ X^{10} $\\
\hline
M $D5$ & $\times$ & $\times$ & $\times$ &$\times$ &$\times$ & $\{a_i\}$ & $\times$ & -- & -- & --\\
k $F1$ & $\times$ & $-$ &$-$ &$-$ &$-$ & $\times$ & -- & -- & -- &--\\
N $NS5$ & $ \times $ &$\times$ &$\times$ &$\times$ &$\times$ &$\times$ & -- &-- &-- &--
\end{tabular}
\end{center}

\begin{figure}[here!]
  \centering
	\includegraphics[width=0.6\textwidth]{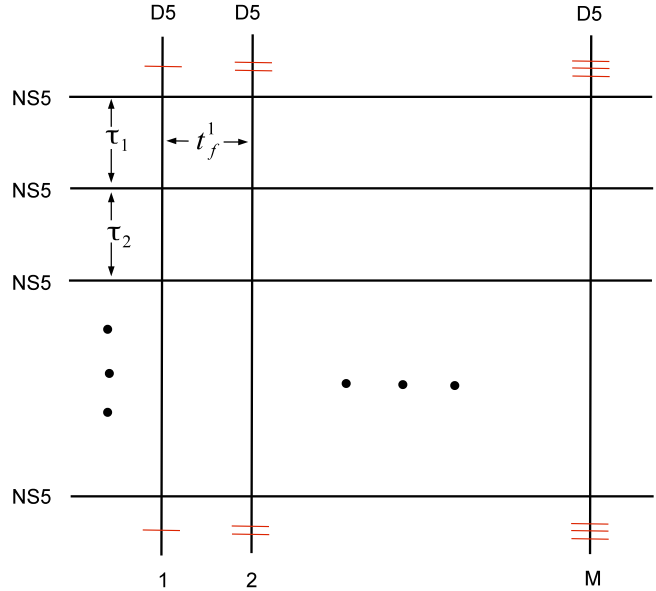}
  \caption{Type IIB brane web.}
  \label{fig:braneweb1}
\end{figure}

\begin{figure}[here!]
  \centering
	\includegraphics[width=0.8\textwidth]{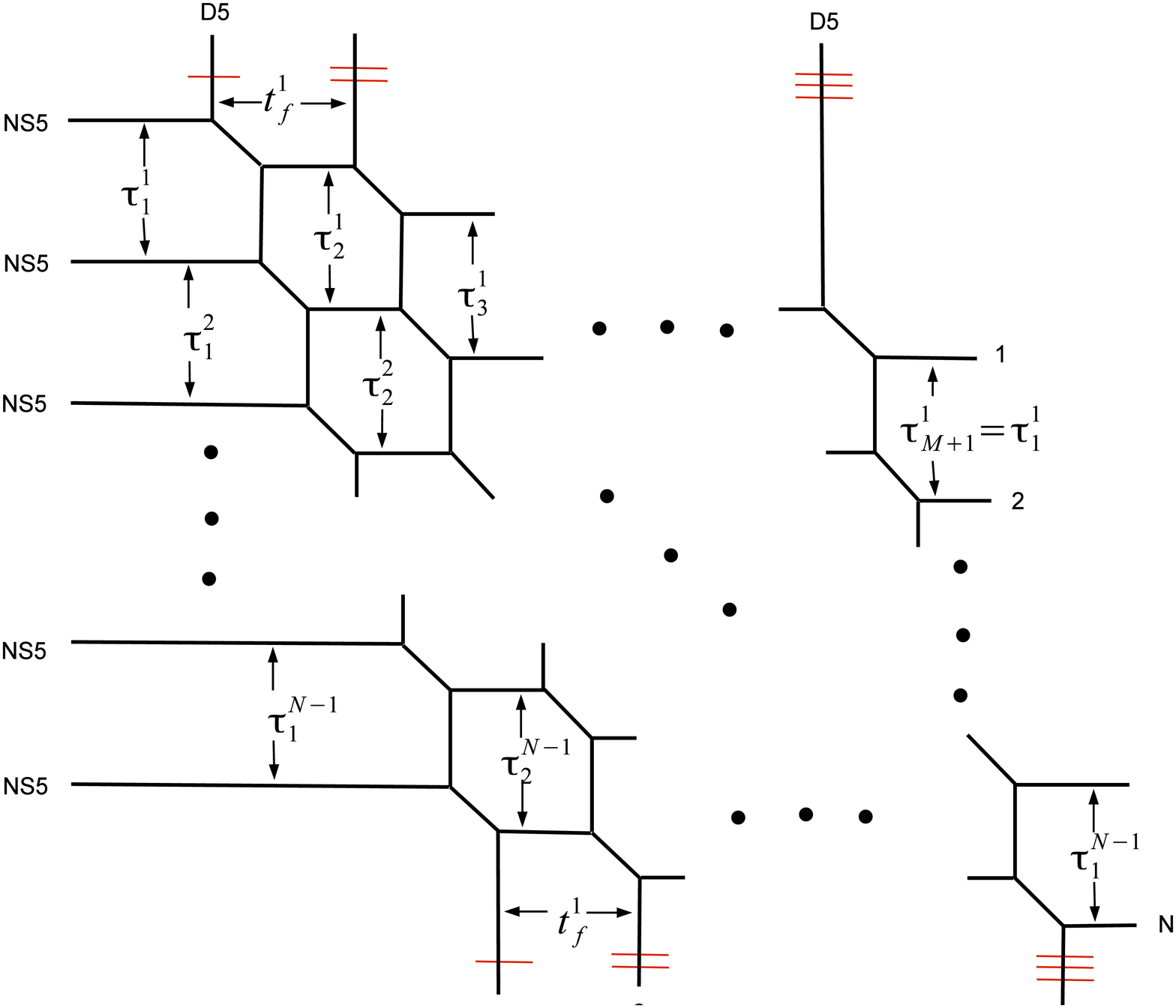}
  \caption{Type IIB brane web with mass deformation.}
  \label{fig:braneweb2}
\end{figure}

Now the $ X^7 $ radius is $ 1/\lambda $, the inverse of the asymptotic radius of the $ \textrm{TN}_N $ circle. It is argued in \cite{Witten:2009xu} that the integral of the B-field on $ C_i $,
\[ \int_{C_i}B = \tau_i,\]
translates after T-duality to the separation between the NS5-branes along the $ X^7 $ direction. This is still valid in the singular limit we are considering where the centers of $ \textrm{TN}_N $ are brought together while leaving the B-flux finite. The D4-branes translate on the type IIB side to D5-branes wrapping the $ X^7 $ circle and sitting at the origin of $ \mathbb{R}^3 $. The resulting brane picture is depicted in Figure \ref{fig:braneweb1}. 
This brane picture describes the subset of parameters in the gauge theory where the Cartan expectation values for all $SU(M)$ gauge factors are the same and the mass is set to zero. This corresponds to a $N + M-1$ dimensional subspace of the full parameter space. To get the full picture after turning on non-zero mass one has to introduce $(1,1)$ branes. These will connect D5-branes which end on NS5-branes from different sides as shown in Figure \ref{fig:braneweb2}. The most general setup of $ (p,q) $-branes now depends on $NM + 1$ parameters and thus reproduces correctly the gauge theory counting.

We complete this chain of dualities by simply recalling the picture of \cite{Leung:1997tw}: type IIB theory on $ S^1 $ (which we later take to be the $X^0$ circle) is the same as M-theory on $ T^2 $, namely a $ (p,q) $-brane corresponds to the $ (p,q) $-cycle of the M-theory $ T^2 $ shrinking over the $ (X^6, X^7) $ base. This way the brane picture uplifts in M-theory to a non-compact Calabi-Yau which is elliptically fibered. For our specific brane setup it turns out that the elliptic fiber is singular and of type $I_N$ in the Kodaira classification of elliptic fibrations \cite{Kodaira}. 
The K\"ahler class $t^M_e$ of the elliptic fiber is identified with the overall gauge coupling of the 5d quiver gauge theory and is thus the inverse of the radius of the $X^1$ circle. That is we have
\begin{equation}
	t^M_e = \frac{1}{R_1}.
\end{equation}
Resolving the singularities of the elliptic fiber leads to various moduli which are identified with the Coulomb branch and mass parameters of the gauge theory.

\subsubsection{Dual gauge theory description}
\label{sec:6d}
Here we will derive a dual six dimensional gauge theory description of our original M-theory setup. To this end 
we start by compactifying on the Taub-NUT circle and pass to the following type IIA description:
\begin{center}
\begin{tabular}{c|cccccccccc}
& $ \mathbb{R} $ & $ S^1 $ & \multicolumn{4}{c}{$\mathbb{R}^4_\parallel$}&$\mathbb{R}$ &\multicolumn{3}{c}{$\mathbb{R}^3_\perp$}\\
& $ X^0 $& $ X^1 $& $ X^2 $& $ X^3 $& $ X^4 $& $ X^5 $& $ X^6 $& $ X^7 $& $ X^8 $& $ X^9 $\\
\hline
M $NS5$ & $\times$ & $\times$ & $\times$ &$\times$ &$\times$ &$\times$ & $\{a_i\}$ & -- & -- & --\\
k $D2$ & $\times$ & $\times$ & $-$ &$-$ &$-$ &$-$ & $\times$ & -- & -- & --\\
N $D6$ &$\times$ &$\times$ &$\times$ &$\times$ &$\times$ &$\times$ &$\times$ &-- &-- &--
\end{tabular}
\end{center}
The centers of Taub-NUT have become D6-branes, the M5-branes have become NS5-branes, and the M2-branes have become D2-branes. The separation between M5-branes simply becomes separation between the NS5-branes. The $ \tau $ parameter is the inverse of the size of the $ X^1 $ circle, multiplied by the radius $\lambda$ of the Taub-NUT circle.

Now we can find out what happens if we perform T-duality along $ X^6 $ (which from now on we must assume to be a circle). The configuration of the branes is as follows:
\begin{center}
\begin{tabular}{c|cccccccccc}
& $ \mathbb{R} $ & $ S^1 $ & \multicolumn{4}{c}{$\mathbb{R}^4$}&\multicolumn{4}{c}{$\textrm{TN}_M$}\\
& $ X^0 $& $ X^1 $& $ X^2 $& $ X^3 $& $ X^4 $& $ X^5 $& $ X^6 $& $ X^7 $& $ X^8 $& $ X^9 $\\
\hline
k $D1$ & $\times$ & $\times$ & $-$ &$-$ &$-$ &$-$ & --& -- & -- & --\\
N $D5$ &$\times$ &$\times$ &$\times$ &$\times$ &$\times$ &$\times$ & -- &-- &-- &--
\end{tabular}
\end{center}
In other words, the $M$ NS5-branes of type IIA in this picture have become $ \textrm{TN}_M $ geometry and the D6-branes have become D5-branes.

\begin{figure}[here!]
  \centering
	\includegraphics[width=0.5\textwidth]{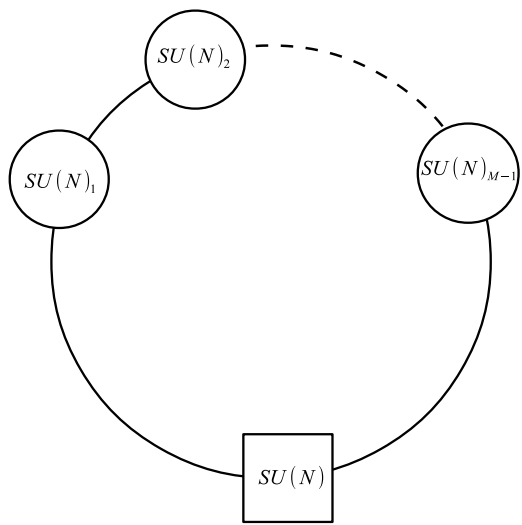}
  \caption{Dual six-dimensional quiver gauge theory.}
  \label{fig:dualgaugeth}
\end{figure}

The theory living on the D5-branes has again an interpretation of a quiver gauge theory. This time, however, each node of the quiver is an $SU(N)$ gauge group with bifundamental matter between adjacent nodes \cite{Douglas:1996sw,Blum:1997mm}. This is depicted in Figure \ref{fig:dualgaugeth}.
As explained in \cite{Blum:1997mm} this gauge theory comes with $M-1$ tensor multiplets and a global $SU(N)$ symmetry. Counting parameters we find $M(N-1)$ Coulomb branch parameters and fugacities which together with the tensor multiplet scalars, the mass parameter and the radius of compactification from six dimensionals to five give $M(N-1) + M-1 + 1 + 1 = M N + 1$ parameters. This matches with the countings from the dual five-dimensional gauge theory and the toric diagram. 

\subsection{Compactification on $S^1 \times S^1$ and relation with topological strings}
\label{sec:s1s1}
Going back to our original M-theory setup, we can also further compactify $X^0$ on $S^1$. While doing this we can introduce the $\Omega$-background by fibering the space $\mathbb{R}^4_{||}$ over this circle. In order to preserve supersymmetry we then also have to fiber $\textrm{TN}_N$ around this circle. Altogether we twist $\textrm{TN}_N\times \mathbb{R}^4_{||}$ by the action of $U(1)\times U(1)$ as we go around the circle parametrized by $X^0$:
\begin{eqnarray}
	U(1)_{\epsilon_1} \times U(1)_{\epsilon_2} & : & (z_1,z_2) \mapsto (e^{2\pi i {\epsilon_1}} z_1, e^{2\pi i \epsilon_2} z_2), \nonumber \\
	~ & : & (w_1,w_2) \mapsto (e^{-\frac{\epsilon_1+\epsilon_2}{2} }w_1,e^{-\frac{\epsilon_1+\epsilon_2}{2}} w_2)
\end{eqnarray}
The second $U(1)$ is nothing else than the isometry $U(1)_b$ of TN$_N$. 

\begin{figure}[here!]
  \centering
  \subfloat[]
{\label{figMAstring}\includegraphics[width=0.5\textwidth]{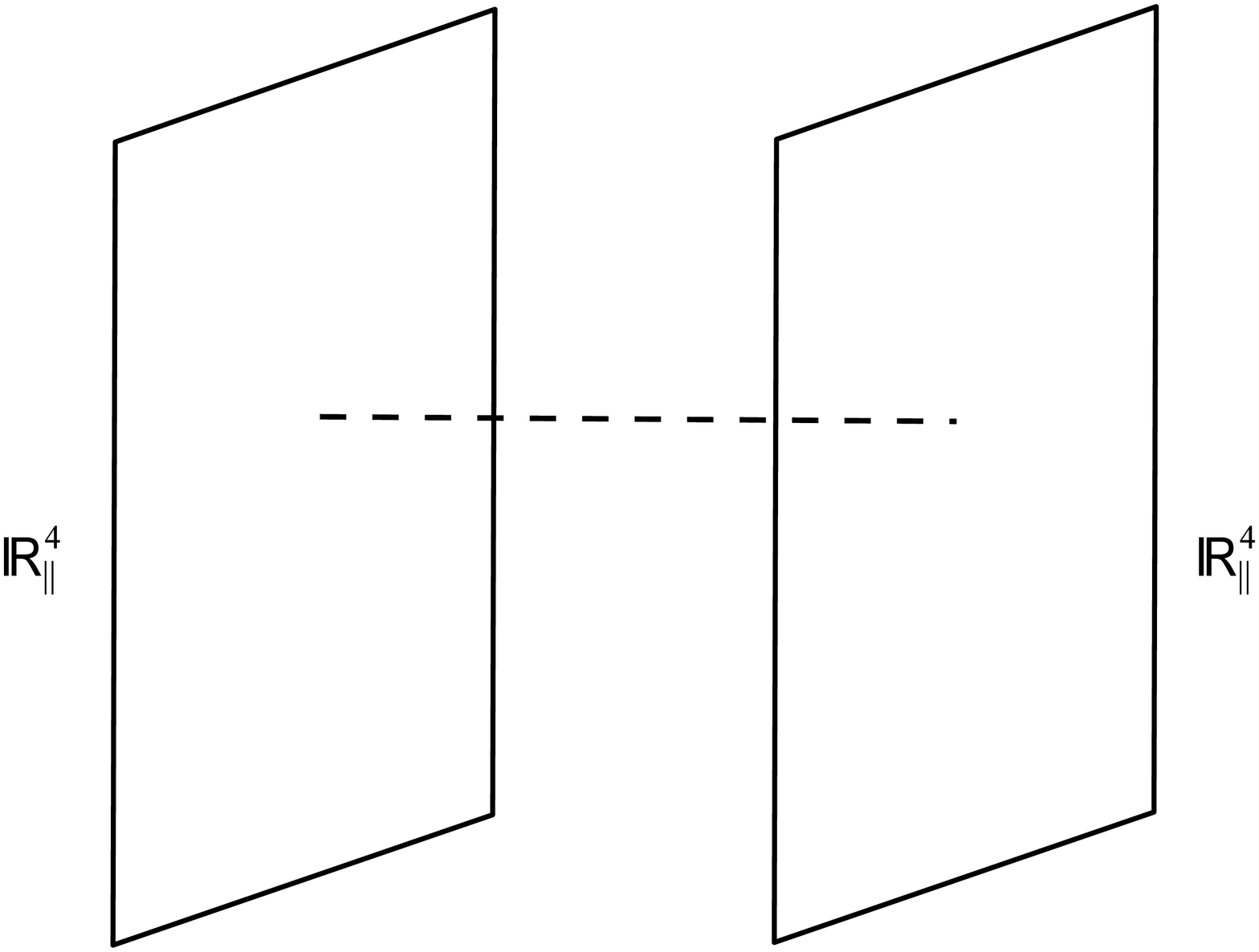}}
  \subfloat[]
{\label{fig:Mstring}\includegraphics[width=0.5\textwidth]{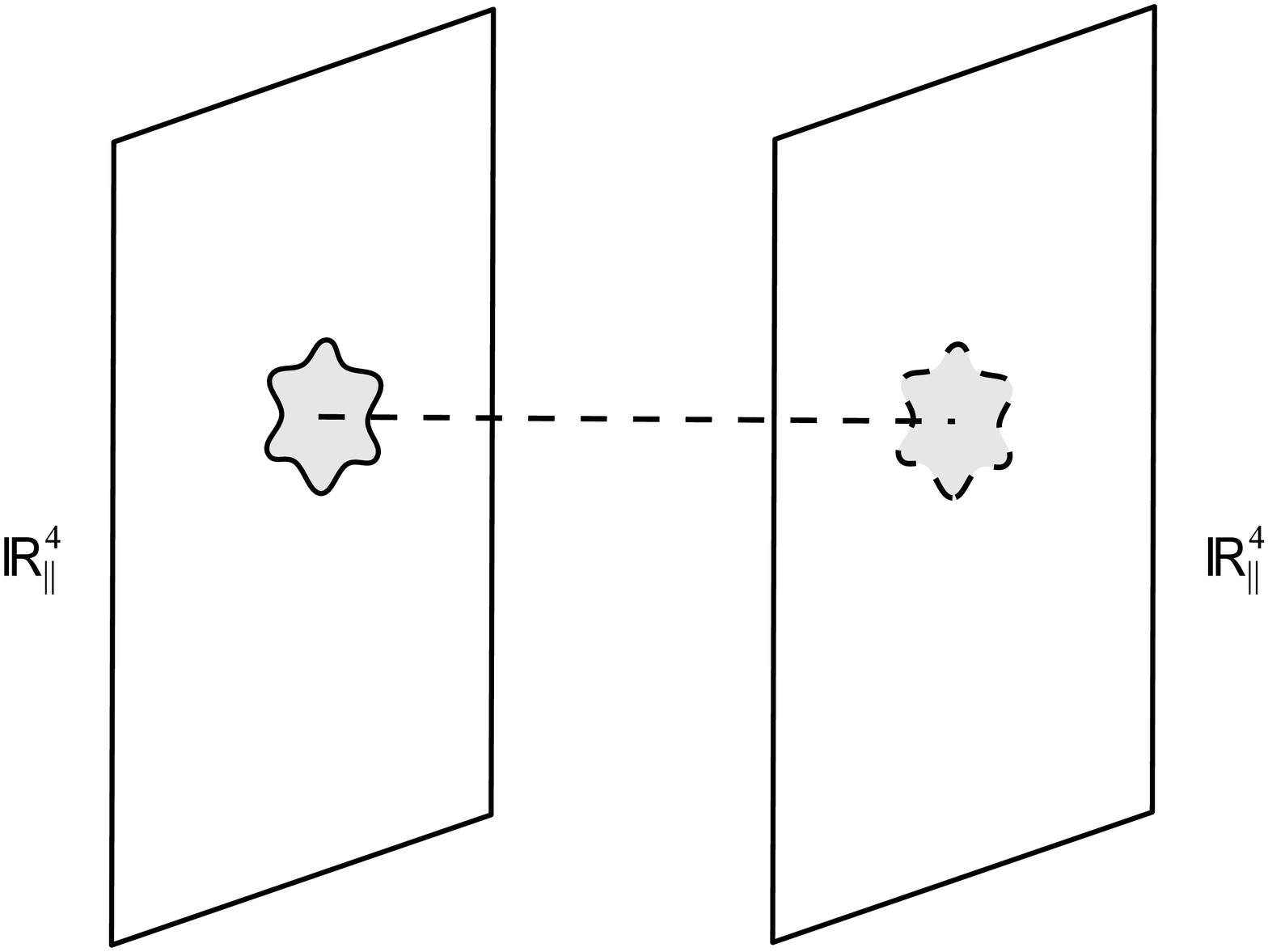}}
  \caption{M-strings versus M$_{A}$-strings. In (a) the gauge group is $U(1)$ and the corresponding
  instantons originating from stretched M2-branes have zero size in the $ \mathbb{R}^4_\parallel $ directions. In (b) we see a thickening of the M2-brane ending on the M5-brane in the case of transverse $A_{N-1}$ singularity, because instantons
  can now acquire a finite size.}
  \label{fig:thickness}
\end{figure}

We can now ask what the theory of the suspended M2-branes is when wrapped around the $X^0$ and $X^1$ directions. A sigma model description can be deduced as follows. The M2-branes as well as the M5-branes will be sitting at the fixed point of the orbifold action in $\mathbb{R}^4_{\perp}$ and as in the M-string setup the M5-branes are extended along $T^2 \times \mathbb{R}^4_{||}$. Also, the M2-branes will appear point-like in $\mathbb{R}^4_{||}$. However, this time their moduli space will not be the one of $U(1)$ instantons but rather that of $SU(N)$ instantons. One way to see this is from the dual type IIB setup described in Section \ref{sec:6d}. From the type IIB brane setup one can see that the D1-branes are instantons from the point of view of the theory living on the D5-branes. As the D1-branes are connected to the M-strings through a chain of dualities we thus see that the moduli space of $k$ $\textrm{M}_{\textrm{A}}$-strings is that of $k$ $SU(N)$ instantons. Furthermore, as the real dimension of this moduli space is $4kN$ we thus see that the $\textrm{M}_{\textrm{A}}$-string has gained more degrees of freedom compared to the M-string whose moduli were the coordinates of $\mathbb{R}^4_{||}$. From another point of view one can say that while the M-string was a point-like object on $\mathbb{R}^4_{||}$ the $\textrm{M}_{\textrm{A}}$-string now fills an extended region in $\mathbb{R}^4_{||}$ because, unlike the $U(1)$ case, the instantons can now
acquire a finite size. Yet from another viewpoint one can say that in the presence of transversal $\textrm{A}_{N-1}$ singularity M2-branes suspended between M5-branes gain thickness (see Figure \ref{fig:thickness}).

The task of the following sections will be to compute these degeneracies and obtain a closed formula for them in terms of the refined topological string partition function.
Again, the partition function of M-theory in this background is by definition the partition function
of the refined topological string on the corresponding Calabi-Yau threefold which now takes the following form:
\bea
Z^{M-theory}((A_{n-1} \times \underbrace{\mathbb{R}^4_{||}) \ltimes T^2_{\epsilon_1,\epsilon_2,m}}_{\mbox{\normalsize{$M {\rm M5}$}}}\times \mathbb{R})=Z_{top}^{refined}(\epsilon_1,\epsilon_2)(CY_{N,M,m,t_f^i,\tau^i_j}).
\eea
As a main tool we will use the topological vertex and its refinement \cite{Aganagic:2003db,Gopakumar:1998ii,Gopakumar:1998jq,Hollowood:2003cv,Iqbal:2007ii}, to compute the degeneracy of BPS states. This correspondence will be used to further extract the elliptic genus of $\textrm{M}_{\textrm{A}}$-strings. These will arise from M2-branes which wrap the torus $T^2$ and are extended along the $X^6$ direction. Having compactified on the second $S^1$ all M-theory parameters get rescaled by the radius $R_0$ and also get complexified due to Wilson lines along the second circle. In particular, by abuse of notation we will now denote the complex structure of $T^2$ by $\tau$. An $\textrm{M}_{\textrm{A}}$-string which has Kaluza-Klein momentum $k$ along the M-theory circle then gives rise to BPS degeneracies which will appear as the coefficient of the $k$-th power of $Q_{\tau} = e^{2\pi i \tau}$ in the topological string partition function of the elliptic Calabi-Yau. Furthermore, such strings can have non-trivial charge under all remaining gauge theory parameters. Their degeneracies appear in the free energy of the topological string as computed in Section \ref{sec:computation}.

\subsection{Quiver theory for the $\textrm{M}_{\textrm{A}}$-strings}
\label{sec:quiver}
A 2d quiver description for the $\textrm{M}_{\textrm{A}}$-strings can be deduced from the type IIB brane setup of $N$ D5-branes probing an $\textrm{A}_{M-1}$ singularity described in Section \ref{sec:6d}. Following \cite{Okuyama:2005gq} the quiver can be constructed from an orbifold of the theory on the D1-branes. Before the orbifolding the theory living on the D1-branes is a $\mathcal{N}=(4,4)$ $U(k)$ gauge theory with one adjoint and $N$ fundamental hypermultiplets. The adjoint hypermultiplet arises from the the $1-1$ strings and the $N$ hypermultiplets come from the $1-5$ strings. To be more specific we have, following \cite{Douglas:1996uz}, the following massless modes on the worldvolume:
\begin{equation}
\begin{array}{cc}
	\textrm{bosons} & \textrm{fermions} \\
	b^{AY} & \psi_-^{A'Y} \\
	b^{A'\tilde{A'}} & \psi_-^{A\tilde{A}'} \\
	A_{--},A_{++} & \psi^{AA'}_+, \psi^{\tilde{A'}Y}_+ \\
	H^{A'} & \chi_-^A, \chi_+^Y,
\end{array}
\end{equation}
where $A_{\pm\pm} = A_0 \pm A_1$.
Furthermore, the indices $(A',\tilde{A'})$ represent the fundamental representations of the two $SU(2)$ groups rotating the directions $X^2,X^3,X^4,X^5$ while $(A,Y)$ are indices for the $SU(2)$'s rotating $X^6,X^7,X^8,X^9$. The scalars in the adjoint $\mathcal{N}=(4,4)$  hypermultiplet are parametrized by $b^{A'\tilde{A'}}$ while those of the vector multiplet are $b^{AY}$. The scalars of the fundamental hypermultiplets, $H^{A'}$, are doublets under $SU(2)_R \equiv SU(2)_{A'}$. The multiplet structure is then obtained by the action of the left-moving and right-moving supercharges:
\begin{equation} \label{eq:susytrafo}
	Q_-^{AA'} b_{A}^{~Y} = \psi_-^{A'Y}, \quad Q_+^{A'Y} b^{~A}_Y = \psi_+^{AA'}.
\end{equation}
These fields can equally well be described in the language of $\mathcal{N}=(2,2)$ chiral and twisted chiral superfields. In particular, the vector multiplet is given by the pair of superfields $(\Sigma,\Phi)$ where $\Sigma$ is a twisted chiral superfield and $\Phi$ is a chiral superfield. Furthermore, the adjoint hypermultiplet is given by the pair of chiral superfields $(B,\tilde{B})$ whereas the fundamental hypermultiplets are $(Q,\tilde{Q})$. That is, we have the following correspondence 
\begin{equation} \label{eq:fields}
	b^{AY} \leftrightarrow (\Sigma,\Phi), \quad b^{A'\tilde{A'}} \leftrightarrow (B,\tilde{B}), \quad H^{A'} \leftrightarrow (Q,\tilde{Q}).
\end{equation}
We next consider orbifolding this theory by $\mathbb{Z}_M$. To preserve the left-moving supersymmetry and break the right-moving one we embed the orbifold group $\mathbb{Z}_M$ in $SU(2)_Y$ giving the following action on fields with $Y$-index
\begin{equation}
	(b^{AY}, \psi_-^{\tilde{A'}Y}, \psi_+^{\tilde{A'}Y}, \chi_+^Y) \mapsto (\zeta^Y b^{AY},\zeta^Y \psi_-^{A'Y},\zeta^Y \psi_+^{\tilde{A'}Y},\zeta^Y \chi_+^Y),
\end{equation}
where $\zeta= e^{\frac{2\pi i}{M}}$ and $Y=\pm$. Note that the remaining fields are invariant under the orbifold action. The resulting theory has $N=(4,0)$ supersymmetry and its field content can equally well be described in the language of $\mathcal{N}=(2,0)$ superfields by decomposing the $\mathcal{N}=(2,2)$ superfields as follows:
\begin{eqnarray}
	\Sigma_{(2,2)}(\theta^+,\bar{\theta}^-) \sim \Sigma - \sqrt{2} \bar{\theta}^+ \Upsilon, & \quad &
	\Phi_{(2,2)}(\theta^+,\bar{\theta}^-) \sim \Phi + \sqrt{2} \theta^+ \Lambda^{\Phi}, \nonumber \\
	B_{(2,2)}(\theta^+,\bar{\theta}^-) \sim B + \sqrt{2} \theta^+ \Lambda^{B}, & \quad &
	\tilde{B}_{(2,2)}(\theta^+,\bar{\theta}^-) \sim  \tilde{B} + \sqrt{2}\theta^+ \Lambda^{\tilde{B}}, \nonumber \\
	Q_{(2,2)}(\theta^+,\bar{\theta}^-) \sim Q + \sqrt{2} \theta^+ \Lambda^Q, & \quad & \tilde{Q}_{(2,2)}(\theta^+,\bar{\theta}^-) \sim Q + \sqrt{2} \theta^+ \Lambda^{\tilde{Q}},
\end{eqnarray}
where $\Sigma$ and $\Phi$, $B$, $\tilde{B}$, $Q$, $\tilde{Q}$ are $(2,0)$ chiral superfields, $\Upsilon$ is the $(2,0)$ gauge superfield, and $\Lambda^i$ is the Fermi superfield.

\begin{figure}[here!]
  \centering
	\includegraphics[width=0.8\textwidth]{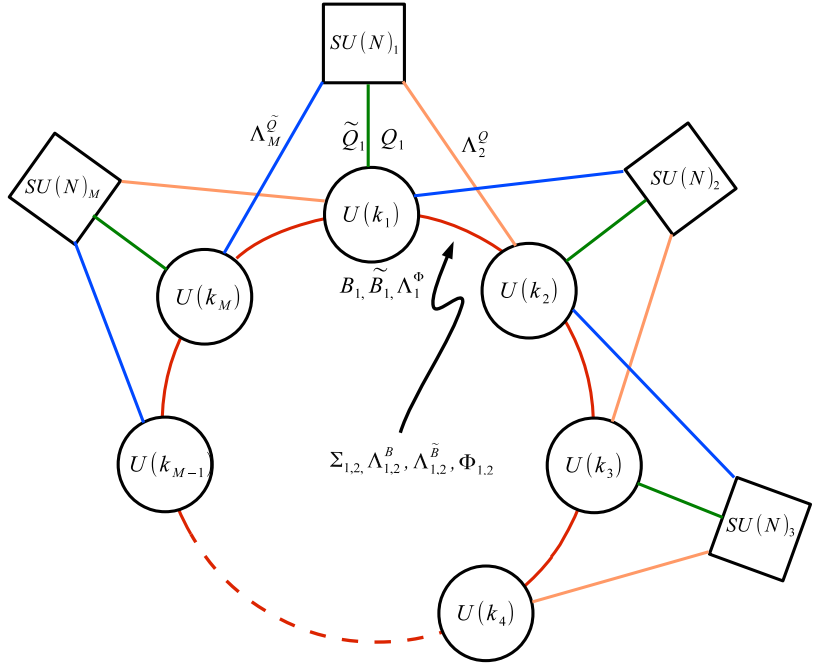}
  \caption{The quiver for the D1-D5-system. In order to obtain the $\textrm{M}_{\textrm{A}}$-strings one has to remove the last node in the inner quiver and all links ending on it. We have also included a representative set of $(2,0)$ fields corresponding to the links connecting the nodes of the quiver.}
  \label{fig:Mstringquiver}
\end{figure}

The orbifolding gives rise to a quiver gauge theory with an inner quiver and an outer one. The inner quiver is the affine $\hat{A}_{M-1}$ Dynkin diagram with nodes corresponding to gauge group factors $U(k_i)$ for $i=1,\cdots,M$ which live on the $i$th copy of D1-branes and are linked by bifundamentals between adjacent nodes. Moreover, there is also an outer $\hat{A}_{M-1}$ quiver with $SU(N)$ nodes which corresponds to the orbifold of the D5-branes. Its nodes are not connected as those modes are not visible from the viewpoint of the D1-branes. However, there are links connecting the outer with the inner quiver. In particular, there are links which connect $SU(N)_i$ nodes of the outer quiver with $U(k_i)$ nodes of the inner one. These links are $(4,0)$ hypermultiplets which are invariant under the $\mathbb{Z}_M$ orbifold action. Matter fields which are not invariant under this action still survive the orbifolding but now reach from $SU(N)_i$ nodes to $U(k_{i-1})$ and $U(k_{i+1})$ nodes. The result is the quiver depicted in Figure \ref{fig:Mstringquiver}.

In order to connect this picture to $\textrm{M}_{\textrm{A}}$-strings we need to turn off D1-brane charge and instead introduce D3-branes wrapped around blow-up cycles of the resolved $A_{M-1}$ singularity. As explained in \cite{Haghighat:2013gba} in type IIB the tension of strings arising from D3-branes wrapping blow-up cycle $C_i$ is given by $t_i = \mu_i/g_s$ where $\mu_i$ is the size of the 2-cycle $C_i$. Taking the limit $\mu_i \rightarrow 0$ with $g_s \rightarrow 0$ decouples the D1-branes and one is left with the D3-branes. In the language of the above quiver this limit corresponds to removing the last node of the inner quiver (i.e. setting its rank to zero) and also all links ending on it. 

\begin{figure}[here!]
  \centering
	\includegraphics[width=0.6\textwidth]{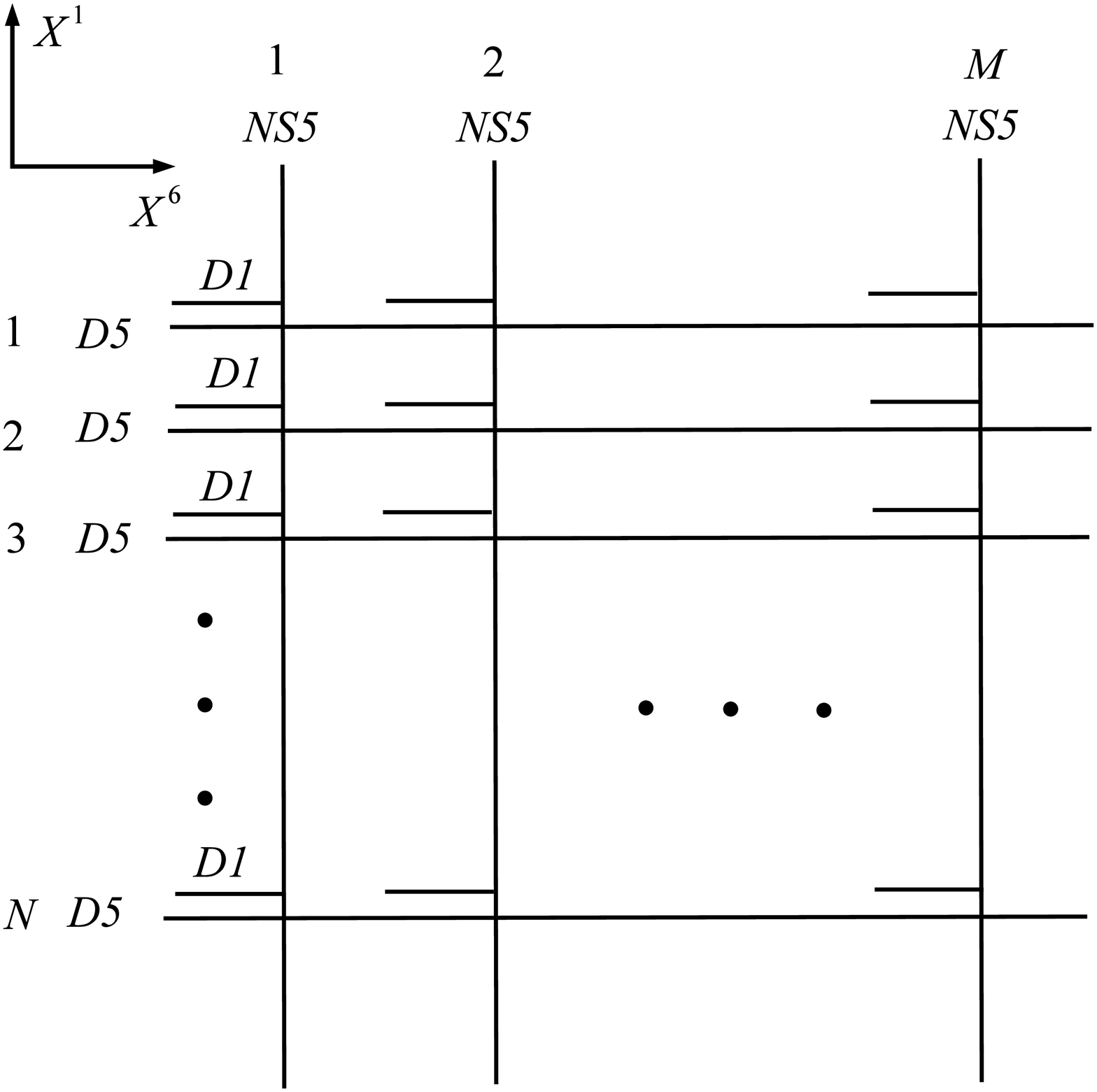}
  \caption{Type IIB brane setup with $M$ NS5-branes and $N$ D5-branes. The D1-branes are parallel to the D5-branes but drawn shorter to distinguish them from the latter ones.}
  \label{fig:quiverbrane1}
\end{figure}

One of the goals of this paper is to make a prediction for the elliptic genus of this quiver using the refined topological vertex which we will put to work in Section \ref{sec:computation}. For this we need to connect the global $U(1)$ symmetries of the quiver to the ones of the original M-theory picture. In particular, we need to identify the mass-rotation $U(1)_m$ as well as the symmetries of the $\Omega$-background, namely $U(1)_{\epsilon_1}$ and $U(1)_{\epsilon_2}$, as a subset of the symmetries of the quiver theory. To this end, it turns out to be useful to study a yet another dual brane setup which captures the field content of the quiver in a very intuitive manner. We start by recalling the type IIA brane setup of Section \ref{sec:6d}:
\begin{center}
\begin{tabular}{c|cccccccccc}
& $ S^1 $ & $ S^1 $ & \multicolumn{4}{c}{$\mathbb{R}^4_\parallel$}&$\mathbb{R}$ &\multicolumn{3}{c}{$\mathbb{R}^3_\perp$}\\
& $ X^0 $& $ X^1 $& $ X^2 $& $ X^3 $& $ X^4 $& $ X^5 $& $ X^6 $& $ X^7 $& $ X^8 $& $ X^9 $\\
\hline
M $NS5$ & $\times$ & $\times$ & $\times$ &$\times$ &$\times$ &$\times$ & $\{a_i\}$ & -- & -- & --\\
k $D2$ & $\times$ & $\times$ & $-$ &$-$ &$-$ &$-$ & $\times$ & -- & -- & --\\
N $D6$ &$\times$ &$\times$ &$\times$ &$\times$ &$\times$ &$\times$ &$\times$ &-- &-- &--
\end{tabular}
\end{center}
Now perform T-duality along the circle in the $X^1$ direction. The result is the type IIB brane setup shown in the table below and is presented pictorially in Figure \ref{fig:quiverbrane1}.
\begin{table}[here!]
\begin{center}
\begin{tabular}{c|cccccccccc}
& $ S^1 $ & $ S^1 $ & \multicolumn{4}{c}{$\mathbb{R}^4_\parallel$}&$\mathbb{R}$ &\multicolumn{3}{c}{$\mathbb{R}^3_\perp$}\\
& $ X^0 $& $ X^1 $& $ X^2 $& $ X^3 $& $ X^4 $& $ X^5 $& $ X^6 $& $ X^7 $& $ X^8 $& $ X^9 $\\
\hline
M $NS5$ & $\times$ & $\times$ & $\times$ &$\times$ &$\times$ &$\times$ & $\{a_i\}$ & -- & -- & --\\
k $D1$ & $\times$ & $-$ & $-$ &$-$ &$-$ &$-$ & $\times$ & -- & -- & --\\
N $D5$ &$\times$ &$-$ &$\times$ &$\times$ &$\times$ &$\times$ &$\times$ &-- &-- &--
\end{tabular}
\end{center}
\end{table}%

\begin{figure}[here!]
  \centering
	\includegraphics[width=0.8\textwidth]{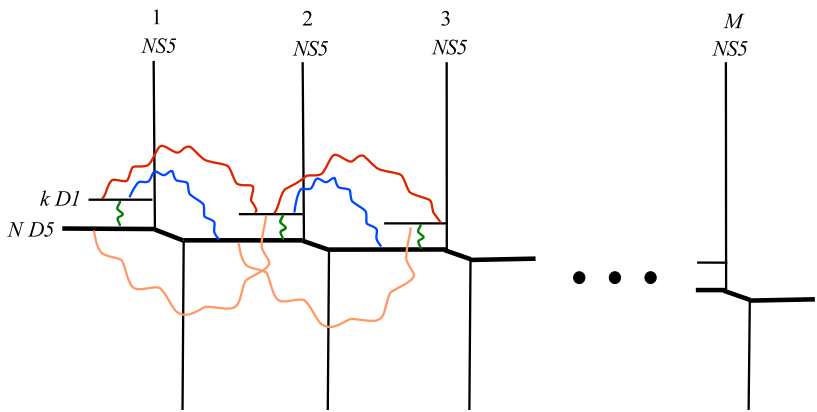}
  \caption{Type IIB brane setup after putting all D5-branes on top of each other. The theory living on the D1-branes corresponds to the quiver gauge theory discussed above. }
  \label{fig:quiverbrane2}
\end{figure}

Note that we have mapped the M-strings to D1-branes which are extended along the $X^6$ direction and wrap the circle parametrized by $X^0$ inside the NS5- and D5-branes. Taking the size of this circle to be very small we can reduce the theory living on the D1-branes along it and the resulting theory is a quantum mechanics living on the segments parametrized by $X^6$. It is now easy to show that the corresponding quiver diagram for this quantum mechanics is exactly the same as the one obtained from the orbifold of the D1-D5 system, depicted in Figure \ref{fig:Mstringquiver}. To see this consider taking all D5-branes to be lying on top of each other so that the strings living on the D1-branes enjoy a full $U(k)$ gauge symmetry and $SU(N)$ flavor symmetry. Furthermore, deform the system by introducing $(N,1)$-branes connecting the D5-branes ending from different sides on the same NS5-brane. The result is depicted in Figure \ref{fig:quiverbrane2}. 
Now one just has to look at the fundamental strings streching between the D1-branes and also the ones ending on the D5-branes. One easily sees that they correspond to the links of the quiver diagram where for ease of identification we have colored the links as well as the strings.

Let us next come to the identification of the global $U(1)$ symmetries. Looking at Figure \ref{fig:quiverbrane2} we can identify the length of the $(N,1)$-branes with the mass-parameter $m$ of the M-theory setup. We can also see that the only strings acquiring mass are the ones reaching from one set of D1-branes to the neighbouring set of either D1- or D5-branes. That is, in the original quiver language the only fields getting massive by turning on non-trivial $m$ are the ones coming from the links connecting nodes of the inner quiver and from links connecting an outer node with adjecent inner nodes. In $(2,0)$ superfield language the first class consists of the twisted chiral multiplets $\Sigma$, the chiral multiplets $\Phi$ and the Fermi superfields $\Lambda^B$ as well as $\Lambda^{\tilde B}$. The second class is formed by the Fermi superfields $\Lambda^Q$ and $\Lambda^{\tilde Q}$. As it is not possible to write down a scalar mass term for these fields in the Lagrangian the mass $m$ has to correspond to the conserved charge of a $U(1)$-current which is a symmetry of the theory. On the other hand from the supersymmetry transformations (\ref{eq:susytrafo}) and the field identifications (\ref{eq:fields}) one can see that the fields $B$, $\tilde B$, $\Lambda^{\Phi}$, $Q$ and $\tilde Q$ carry either an $A'$ or a $\tilde{A'}$ index which shows that they transform nontrivially under rotations of $\mathbb{R}^4_{||}$. They will thus carry $U(1)_{\epsilon_i}$ charge. As a clarifying example and also to set our conventions we give here the charges of the fields under the various $U(1)$'s for the case where $M=2$, that is when the D3-quiver contains only one inner node:
\begin{center}
	\begin{tabular}{|c|c|c|c|c|c|c|c|}
	\hline
		~ & $\Lambda^{\Phi}$ & $B$ & $\tilde{B}$ & $Q$ & $\tilde{Q}$ & $\Lambda^Q$ & $\Lambda^{\tilde Q}$ \\
		\hline
		$U(k)$ & $\textrm{adj}$ & $\textrm{adj}$ & $\textrm{adj}$ & $\square$ & $\overline{\square}$ & $\square$ & $\overline{\square}$\\
		\hline
		$U(1)_{\epsilon_1}$ & $-1$ & $1$ & $0$ & $\frac{1}{2}$ & $\frac{1}{2}$ & $0$ & $0$ 	\\
		\hline
		$U(1)_{\epsilon_2}$ & $-1$ & $0$ & $1$ & $\frac{1}{2}$ & $\frac{1}{2}$ & $0$ & $0$\\
		\hline
		$U(1)_m$ & $0$ & $0$ & $0$ & $0$ & $0$ & $1$ & $1$\\
		\hline
	\end{tabular}
\end{center}
 
Let us next comment on the Higgs branch of the quiver gauge theory.  As the $(4,0)$-theory contains superpotential terms coming from the faces of the quiver we have to restrict the parameters of this superpotential in order to make contact with the M-theory setup. We claim that the answer for the elliptic genus of this quiver gauge theory computed along the lines of \cite{Benini:2013nda,Gadde:2013dda,Benini:2013xpa} matches the topological vertex result of Section \ref{sec:computation} where $m$ controls the ``mass-rotation'' of the M-theory setup. We will return to this point in Section \ref{sec:eg}, where we will be able to perform an explicit check in the case of $ M = 2 $, $ N=1 $. Indeed as argued in \cite{Okuyama:2005gq} the Higgs branch moduli space of the quiver consists of $M-1$ copies of the moduli space of $k$ $SU(N)$ instantons and hence the quiver also contains the sigma model description for the $\textrm{M}_\textrm{A}$-strings. The bosons of the sigma model will arise from the $4kN$ bosonic zero modes in the $k$ $SU(N)$ instanton background and are thus sections of the tangent bundle. Left-moving fermions are again sections of the tangent bundle whereas right-moving fermions transform as sections of a different bundle breaking supersymmetry in the right-moving sector. In the case where the inner quiver contains only one node, that is where there are only two M5-branes in the M-theory setup, this bundle is formed by the $2kN$ fermionic zero modes of the Dirac equation for an adjoint fermion in the instanton background together with their complex conjugates. For more details on this bundle and its Chern characters we refer to \cite{Losev:2003py}.

For the general quiver with $M-1$ nodes  the picture is more complicated. The bosons are sections of the tangent bundle of the moduli space
\begin{equation}	\label{eq:instmoduli}
	\mathcal{M}^N_{k_1,k_2,\cdots,k_{M-1}} \equiv \mathcal{M}(k_1,N)\times\mathcal{M}(k_2,N)\times \cdots \times \mathcal{M}(k_{M-1},N).
\end{equation}
The right-moving fermions are sections of a bundle $V$ which is of same dimensionality as the tangent bundle. It admits a decomposition
\begin{equation}
	V = \bigoplus_{s=0}^{M-1} V_s,
\end{equation}
where the $V_s$ are bundles over $\mathcal{M}(k_s,N)\times \mathcal{M}(k_{s+1},N)$ and it is understood that $\mathcal{M}(k_0,N)$ and $\mathcal{M}(k_M,N)$ are empty spaces. The moduli space of $k_s$ instantons in $SU(N)$ gauge theory admits fixed points under the $U(1)_{\epsilon_1} \times U(1)_{\epsilon_2}\times U(1)^N$ action on ADHM data which are themselves labelled by ADHM data for an $N$-tuple of $U(1)$ instantons: $(k_s^1,k_s^2,\cdots,k_2^N)$ with the property $\sum_{a=1}^N k_s^a = k_s$. The moduli space of $U(1)$ instantons is the Hilbert scheme of points on $\mathbb{C}^2$ and fixed points on  $\textrm{Hilb}^{k_s^i}(\mathbb{C}^2)$ are labelled by codimension $k_s^a$ Ideals in $\mathbb{C}[x,y]$ denoted by $I_s^a$. Thus the fixed points on $\mathcal{M}(k_s,N)\times\mathcal{M}(k_{s+1},N)$ can be identified by pairs of ideals and in this language the bundle $V_s$ restricted to these fixed points is of the form
\begin{equation}
	\left.V_s \right|_{\textrm{fixed points}} = \left( \oplus_{a=1,b=1}^{N} \textrm{Ext}^1(I_s^a,I_{s+1}^b)\right) \otimes L^{-\frac{1}{2}},
\end{equation} 
where $L$ is the canonical line bundle on $\mathbb{C}^2$ and $I_0$ and $I_M$ are co-dimension zero Ideals. In Section \ref{sec:eg} we will explain how an explicit description of these bundles gives another way of computing the elliptic genus of $ \mathrm{M}_{\mathrm{A}} $-strings.

\section{Topological string computation of the partition function} 
\label{sec:computation}
The goal of this section is to compute the topological partition function of M5-branes on the geometry $\mathbb{R}^4_{||} \times T^2 \times \mathbb{R} \times A_{N-1}$ presented in Section 2. To this end we compute the refined topological string partition function of the non-compact Calabi-Yau given by the toric diagram in Figure \ref{fig:braneweb2}. In computing such a partition function we have to specify a choice of \textit{preferred direction}, which can be taken either to be the vertical axis or the horizontal axis. Choosing the vertical axis as preferred direction will lead to the Nekrasov partition function for the five-dimensional gauge theory given by the quiver of Figure \ref{fig:5dquiver1} (in line with the duality
frame of type IIA with D4 branes probing $A_{N-1}$ singularity), whereas the choice of the horizontal axis leads to the Nekrasov partition function for the dual six-dimensional gauge theory of Figure \ref{fig:dualgaugeth} (corresponding to the duality frame involving $N$ D5-branes of type IIB
probing $A_{M-1}$ singularity). In order to extract the elliptic genus of $\textrm{M}_\textrm{A}$-strings we have to compute the latter partition function. We do this in steps. First, in Section \ref{sec:strip} we study the holomorphic curves contributing to the open topological string partition function for a certain periodic strip geometry (illustrated in Figure \ref{fig:infinitestrip} in the of case $ N=2 $). In Section \ref{sec:domainwall} we normalize the open topological string partition function for this periodic strip by the contributions of closed topological strings (that is, by the partition function of a single M5-brane on transverse $ A_{N-1} $ singularity). The resulting expression, equation \eqref{eq:ANDomain}, is given an interpretation as a domain wall for the theory of M2-branes on $\mathbb{R}\times T^2$ in presence of a transverse $ A_{N-1} $ singularity. In Section \ref{sec:mapartition} we glue together the contributions from the $ M $ different strips geometries that the toric Calabi-Yau is built out of to obtain the partition function of our system of $ M $ M5-branes on transverse $ A_{N-1} $ singularities, normalized by the $ M $-th power of the partition function of a single $ M5 $-brane, expressed as a sum of $ \mathrm{M}_{\mathrm{A}} $-string contributions. This is the main computational result of our paper, and is given in equation \eqref{eq:partitionfunction}. We also comment on the manifest modular properties of the partition function. Finally, in Section \ref{sec:eg} we discuss other approaches for directly computing the elliptic genus of $ \textrm{M}_A $-strings: either by studying the appropriate bundles over the moduli space of instantons \eqref{eq:instmoduli}, or by computing the 2d index of the (4,0) quiver gauge theory of Section \ref{sec:quiver}.

\subsection{Periodic strip partition function from curve counting}\label{sec:strip}
\par{The relevant geometry to compute the topological string partition function for M-strings at $A_{N-1}$ singularities is the partial compactification of the so-called strip geometry, replacing the resolved conifold geometry of the original M-stings setup. The length of the strip is determined by $N$; more specifically, $ N $ is the number of external legs on each side of the strip. In \cite{Haghighat:2013gba}, the refined topological vertex is adapted to compute the topological string amplitudes. The recursive method used there can be employed in the present case as well; however, the computations get very cumbersome, even for the $A_{1}$ singularity. Instead we follow a more intuitive approach based on an observation of \cite{Iqbal:2008ra}.} 
\par
\begin{figure}[h]
  \begin{center}
    \includegraphics[width=0.8\textwidth]{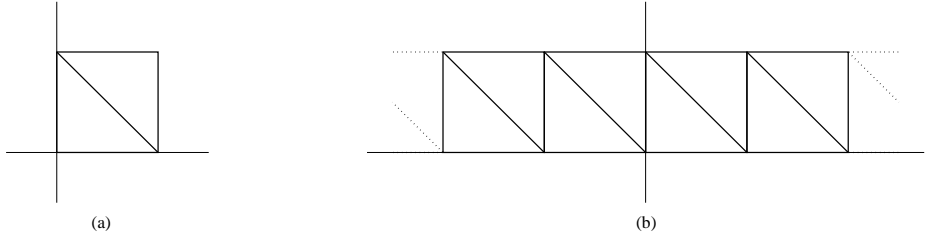}
  \end{center}
  \caption{(a) The Newton polygon for the resolved conifold, and (b) the cover of the Newton polygon after partially compactification of the resolved conifold along the horizontal edges.}\label{newton}
\end{figure}

{Let us briefly review the observation of \cite{Iqbal:2008ra}: the topological partition function for the partial compactification of the resolved conifold can be computed by counting the holomorphic maps in an infinite, but periodic, strip geometry. The Newton polygon of the resolved conifold is depicted in Figure \ref{newton}(a) and is obviously planar. However, the Newton polygon of the partial compactification of the resolved conifold is non-planar and lives on a cylinder. In the covering space of the cylinder it can be represented as a periodic configuration. The holomorphic curves wrap the compact part of the geometry which consist of an infinite chain of ${\mathbb P}^{1}$'s.      }
\par{A holomorphic curve $C$ satisfies $C\cdot C=2g-2$ for $g\geq 0$, where $g$ is the genus of the curve $C$. Two rational curves $C_{1}$ and $C_{2}$ with vanishing intersection, $C_{1}\cdot C_{2}=0$, do not form a holomorphic curve $C_{1}+C_{2}$, since $(C_{1}+C_{2})\cdot(C_{1}+C_{2})=-4$. In other words, if $C_{1}$ and $C_{2}$ are not connected $C_{1}+C_{2}$ is not holomorphic. However, if $C_{1}$ and $C_{2}$ have the intersection number 1, $C_{1}+C_{2}$ is a holomorphic curve of genus zero, since then $(C_{1}+C_{2})\cdot(C_{1}+C_{2})=-2$. From this discussion we can conclude that the individual $\mathbb{P}^{1}$'s and any connected chain of them contribute to the A-model topological string partition function. We need to identify all possible such curves.}
\par
\begin{figure}[h]
  \begin{center}
    \includegraphics[width=0.4\textwidth]{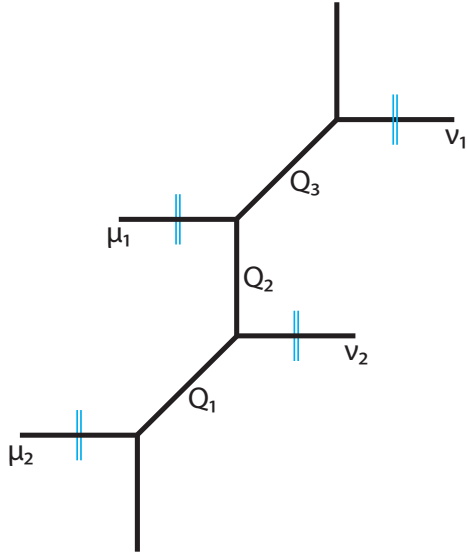}
  \end{center}
  \caption{The basic building block to compute the topological string partition function for the periodic strip. The small double lines (blue) denote the choice of the preferred direction of the refined topological vertex.}\label{twostrip}
\end{figure}

{In our case, the conifold is replaced with the so-called strip geometry and we need to consider this simple building block as one of the periods of the Newton polygon. In contrast to \cite{Iqbal:2008ra}, all the external legs are labelled by Young diagrams; we are interested in constructing the ``domain walls'' for $\mbox{M}_{A}$ strings. It turns out that the detailed understanding of the strip geometry with two external legs is enough to construct the partition function for the infinite strip. The partition function for such a strip, Figure \ref{twostrip}, is given by  }
\begin{align}\label{simplestrip}\nonumber
Z^{\mu_{1}\mu_{2}}_{\nu_{1}\nu_{2}}&=q^{-\frac{\Arrowvert\mu_{1}^{t}\Arrowvert^{2}+\Arrowvert\mu_{2}^{t}\Arrowvert^{2}}{2}}t^{-\frac{\Arrowvert\nu_{1}\Arrowvert^{2}+\Arrowvert\nu_{2}\Arrowvert^{2}}{2}}\widetilde{Z}_{\mu_{1}^{t}}(t^{-1},q^{-1})\widetilde{Z}_{\mu_{2}^{t}}(t^{-1},q^{-1})\widetilde{Z}_{\nu_{1}}(q^{-1},t^{-1})\widetilde{Z}_{\nu_{2}}(q^{-1},t^{-1})\\
&\times\prod_{i,j=1}^{\infty}\frac{(1-Q_{A}\,t^{\mu_{2,i}-j+1/2}q^{\nu_{2,j}^{t}-i+1/2} )(1-Q_{B}\,t^{\nu_{2,i}-j+1/2}q^{\mu_{1,j}^{t}-i+1/2} )}{(1-Q_{A}Q_{B}\,t^{\mu_{2,i}-j}q^{\mu_{1,j}^{t}-i+1})}\nonumber\\
&\times\frac{(1-Q_{C}\,t^{\mu_{1,i}-j+1/2}q^{\nu_{1,j}^{t}-i+1/2} )(1-Q_{A}Q_{B}Q_{C}\,t^{\mu_{2,i}-j+1/2}q^{\nu_{1,j}^{t}-i+1/2} )}{(1-Q_{B}Q_{C}\,t^{\nu_{2,i}-j+1}q^{\nu_{1,j}^{t}-i})},
\end{align}
where we have used $\Arrowvert\mu\Arrowvert^{2}=\sum_{i=1}^{\ell{\mu}}\mu_{i}^{2}$ and the specialisation of the Macdonald polynomial 
\[\widetilde Z_\nu(t^{-1},q^{-1}) = \prod_{(i,j)\in \nu}\left(1-q^{j-\nu_i}t^{i-\nu_j^t-1}\right)^{-1}. \]
\par
\begin{figure}[h]
  \begin{center}
    \includegraphics[width=0.4\textwidth]{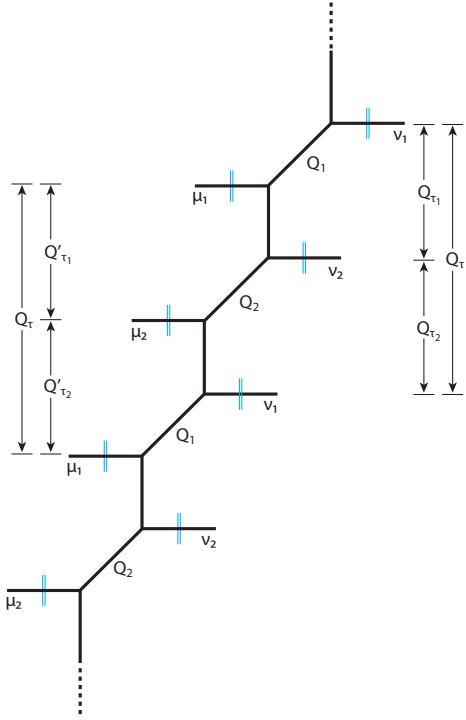}
  \end{center}
  \caption{The (periodic) toric diagram with a basic strip of ``length'' $N=2$ used in computing the M5-brane partition function in the presence of a transverse $A_{1}$ singularity. The small double lines (blue) denote the choice of the preferred direction of the refined topological vertex.}\label{fig:infinitestrip}
\end{figure}

{The contributions coming from ${\cal O}(-1)\oplus{\cal O}(-1)\mapsto{\mathbb P}^{1}$ curves are easy to determine and all have the same form. We can easily distinguish between the curves ${\cal O}(-2)\oplus{\cal O}(0)\mapsto{\mathbb P}^{1}$ (labelled by $\mu_{1}$ and $\mu_{2}$) and ${\cal O}(0)\oplus{\cal O}(-2)\mapsto{\mathbb P}^{1}$ (labelled by $\nu_{1}$ and $\nu_{2}$), which is reflected by the different exponents of $q$ and $t$ in the factors above.} 
Before spelling out the partition function relevant for the $A_{N-1}$ singularity, let us demonstrate our derivation for the $A_{1}$ singularity, depicted in Figure \ref{fig:infinitestrip}; the generalization will be obvious. 

\vspace{1cm}
\underline{${\cal O}(-1)\oplus{\cal O}(-1)\mapsto{\mathbb P}^{1}$}
\vspace{0.5cm}
\par{The curves which belong to this class are labelled by $(\mu_{a},\nu_{b})$ or $(\nu_{a},\mu_{b})$\footnote{The first partition is always taken to be lower than the second partition in the toric diagram.}. We need to take into account all holomorphic curves from a partition $\mu_{a} (\nu_{a})$ to another partition $\nu_{b} (\mu_{b})$. Its contribution to the partition function has the following form for $(\mu_{a},\nu_{b})$}
\begin{align}
\prod_{i,j=1}^{\infty}(1-{\cal Q}_{ab}\,t^{\mu_{a,i}-j+1/2}q^{\nu_{b,j}^{t}-i+1/2}),
\end{align}
where ${\cal Q}_{ab}$ denotes the corresponding K\"{a}hler parameter of the class spanned between the two partitions; we will later give explicit expressions for them. Note that we get infinitely many contributions since the strip is periodic for each pair partitions $(\mu_{a},\nu_{b})$ and $(\nu_{a},\mu_{b})$. In the case of the $A_{1}$ singularity, we have the following curves contributing

\begin{equation}
\begin{array}{llll}\label{classes}
(\mu_{1},\nu_{1};Q_{1}),\qquad\,\,&(\mu_{1},\nu_{2};Q_{1}Q_{\tau_{2}}),\qquad\,\,&(\mu_{2},\nu_{1};Q_{2}Q_{\tau_{1}}),\qquad\,\,&(\mu_{2},\nu_{2};Q_{2}),\\
(\nu_{1},\mu_{1};Q_{1}^{-1}Q_{\tau}),&(\nu_{1},\mu_{2};Q_{2}^{-1}Q_{\tau_{2}}),&(\nu_{2},\mu_{1};Q_{1}^{-1}Q_{\tau_{1}}),&(\mu_{2},\nu_{2};Q_{2}^{-1}Q_{\tau}),
\end{array} 
\end{equation}
with $Q_{\tau}\equiv Q_{\tau_{1}}Q_{\tau_{2}}$\footnote{Let us make a remark about our notation. In the present case, the geometry possesses more K\"{a}hler parameters than in the case considered in \cite{Haghighat:2013gba}. In that case (where $ N = 1 $) we have $Q_{1}\equiv Q_{m}$, which corresponds to the adjoint mass of the 5d $N=2^{*}$ theory. When $N\neq 1$, the $Q_{i}$'s are indirectly related to the bifundamental hypermultiplet masses. We also have $ N $ parameters $Q_{\tau_{1}},\dots, Q_{\tau_N}$ which are related to the gauge theory coupling constants of the corresponding nodes of the quiver; they satisfy $ Q_{\tau_1}\cdot\ldots\cdot Q_{\tau_N} = Q_\tau.$}, the K\"ahler parameter associated to the elliptic fiber. Therefore we have the following infinite products
\begin{align}\nonumber
&\prod_{k=1}^{\infty}\prod_{i,j=1}^{\infty}(1-Q_{1}Q_{\tau}^{k-1}\,t^{\mu_{1,i}-j+1/2}q^{\nu_{1,j}^{t}-i+1/2})(1-Q_{1}Q_{\tau_{2}}Q_{\tau}^{k-1}\,t^{\mu_{1,i}-j+1/2}q^{\nu_{2,j}^{t}-i+1/2})\\\nonumber
&\times(1-Q_{2}Q_{\tau_{1}}Q_{\tau}^{k-1}\,\,t^{\mu_{2,i}-j+1/2}q^{\nu_{1,j}^{t}-i+1/2})(1-Q_{2}Q_{\tau}^{k-1}\,\,t^{\mu_{2,i}-j+1/2}q^{\nu_{2,j}^{t}-i+1/2})\\\nonumber
&\times(1-Q_{1}^{-1}Q_{\tau}^{k}\,t^{\nu_{1,i}-j+1/2}q^{\mu_{1,j}^{t}-i+1/2})(1-Q_{2}^{-1}Q_{\tau_{2}}Q_{\tau}^{k-1}\,t^{\nu_{1,i}-j+1/2}q^{\mu_{2,j}^{t}-i+1/2})\\
&\times(1-Q_{1}^{-1}Q_{\tau_{1}}Q_{\tau}^{k-1}\,t^{\nu_{2,i}-j+1/2}q^{\mu_{1,j}^{t}-i+1/2})(1-Q_{2}^{-1}Q_{\tau}^{k}\,t^{\nu_{2,i}-j+1/2}q^{\mu_{2,j}^{t}-i+1/2}),
\end{align} 
where we have included the factors $Q_{\tau}^{k-1}$ reflecting the periodicity of the Newton polygon, i.e., all the other curves in addition to the initial ones listed in Eq.(\ref{classes}).

\vspace{1cm}
\underline{${\cal O}(-2)\oplus{\cal O}(0)\mapsto{\mathbb P}^{1}$}
\vspace{0.5cm}
\par{As mentioned before these are the curves labelled by $(\mu_{a},\mu_{b})$ and their contributions to the partition function can be obtained from Eq.(\ref{simplestrip}):}
\begin{align}
\prod_{i,j=1}^{\infty}\frac{1}{(1-{\cal Q}_{ab}\,t^{\mu_{a,i}-j}q^{\mu_{b,j}^{t}-i+1})},
\end{align}
with the appropriate K\"{a}hler factors ${\cal Q}_{ab}$. This class includes the following curves 
\begin{equation}
\begin{array}{llll}
(\mu_{1},\mu_{1};Q_{\tau}),\qquad\,\,&(\mu_{1},\mu_{2};Q'_{\tau_{2}}),\qquad\,\,&(\mu_{2},\mu_{1};Q'_{\tau_{1}}),\qquad\,\,&(\mu_{2},\mu_{2};Q_{\tau}).
\end{array} 
\end{equation}
We can immediately determine the corresponding amplitudes for these curves 
\begin{align}\nonumber
&\prod_{k=1}^{\infty} (1-Q_{\tau}^{k}\,t^{\mu_{1,i}-j}q^{\mu_{1,j}^{t}-i+1})^{-1}(1-Q'_{\tau_{2}}Q_{\tau}^{k-1}\,t^{\mu_{1,i}-j}q^{\mu_{2,j}^{t}-i+1})^{-1}(1-Q'_{\tau_{1}}Q_{\tau}^{k-1}\,t^{\mu_{2,i}-j}q^{\mu_{1,j}^{t}-i+1})^{-1}\\
&\times (1-Q_{\tau}^{k}\,t^{\mu_{2,i}-j}q^{\mu_{2,j}^{t}-i+1})^{-1},
\end{align}
where the inclusion of the $Q_{\tau}$ again reflects the periodicity. 

\vspace{1cm}
\underline{${\cal O}(0)\oplus{\cal O}(-2)\mapsto{\mathbb P}^{1}$}
\vspace{0.5cm}
\par{These curves are labeled by $(\nu_{a},\nu_{b})$ and their contribution is close to the ones coming from ${\cal O}(-2)\oplus{\cal O}(0)\mapsto{\mathbb P}^{1}$ except the changes in the exponents,}
 \begin{align}
\prod_{i,j=1}^{\infty}\frac{1}{(1-{\cal Q}_{ab}\,t^{\nu_{a,i}-j+1}q^{\nu_{b,j}^{t}-i})}.
\end{align}
Let us again list the curves in the present case,
\begin{equation}
\begin{array}{llll}
(\nu_{1},\nu_{1};Q_{\tau}),\qquad\,\,&(\nu_{1},\nu_{2};Q_{\tau_{2}}),\qquad\,\,&(\nu_{2},\nu_{1};Q_{\tau_{1}}),\qquad\,\,&(\nu_{2},\nu_{2};Q_{\tau}).
\end{array} 
\end{equation}
The partition function will also include the following factors:
\begin{align}\nonumber
&\prod_{k=1}^{\infty} (1-Q_{\tau}^{k}\,t^{\nu_{1,i}-j+1}q^{\nu_{1,j}^{t}-i})^{-1}(1-Q_{\tau_{2}}Q_{\tau}^{k-1}\,t^{\nu_{1,i}-j+1}q^{\nu_{2,j}^{t}-i})^{-1}(1-Q_{\tau_{1}}Q_{\tau}^{k-1}\,t^{\nu_{2,i}-j+1}q^{\nu_{1,j}^{t}-i})^{-1}\\
&\times (1-Q_{\tau}^{k}\,t^{\nu_{2,i}-j+1}q^{\nu_{2,j}^{t}-i})^{-1}. 
\end{align}
All these factors we obtain using the above approach match the explicit calculation we performed following the methods of \cite{Haghighat:2013gba} up to a factor of $\eta(\tau)^{-1}$. 

\vspace{1cm}
\underline{$A_{N-1}\,\mbox{Singularity}$}
\vspace{0.5cm}
\par{From the above discussion the generalisation for the $A_{N-1}$ singularity is immediate. The numerator will have the form}
\begin{align}
\prod_{a,b=1}^{N}\prod_{i,j,k=1}^{\infty}(1-Q_{\tau}^{k-1}Q_{ab}\,t^{\mu_{a,i}-j+1/2}q^{\nu_{b,j}^{t}-i+1/2})(1-Q_{\tau}^{k-1}\bar{Q}_{ba}\,t^{\nu_{b,i}-j+1/2}q^{\mu_{a,j}^{t}-i+1/2}),
\end{align}
such that 
\begin{align}\label{id}
Q_{ab}\bar{Q}_{ba}=Q_{\tau},
\end{align}
where we define $Q_{\tau}\equiv\prod_{i=1}^{N}Q_{\tau_{i}}$ for the $A_{N-1}$ singularity. The last equality has a simple explanation: $Q_{ab}$ and $\bar{Q}_{ba}$ are defined on the basic strip, c.f. Figure \ref{fig:infinitestrip}. The geometry we are interested in is the partial compactification of this basic geometry. The parameter $Q_{ab}$ measures the distance between partitions $\mu_{a}$ and $\nu_{b})$, and $\bar{Q}_{ba}$ measures the distance between partitions $\nu_{b}$ and $\mu_{a})$. Together they add up to the circumference of the cylinder the Newton polygon is wrapped on. We will label the K\"{a}hler classes for the $(\nu_{a+1},\nu_{a})$ and $(\mu_{a+1},\mu_{a})$ curves by $Q_{\tau_a}$, for $a=1,\mathellipsis,N-1$. The class for curves $(\nu_{1},\nu_{N})$ and $(\mu_{1},\mu_{N})$ is denoted by $Q_{\tau_N}$ (depicted in Figure \ref{fig:infinitestrip}). With this definition, $Q_{ab}$'s can be written as
\begin{equation}\label{eq:Qab}
Q_{ab}=\left\{ \begin{array}{lll}Q_{a}\prod_{j=b}^{N}Q_{\tau_{j}},&(\mbox{mod}\,Q_{\tau})& \mbox{for}\, a=1,\\ 
                                            Q_{a}\prod_{i=1}^{a-1}Q_{\tau_{i}}\prod_{j=b}^{N}Q_{\tau_{j}},&(\mbox{mod}\,Q_{\tau})&\mbox{for}\, a\neq1, \end{array}\right.
\end{equation}
where $(\mbox{mod}\,Q_{\tau})$ means that any $Q_{\tau}$ appearing in the definition of $Q_{ab}$ is set to 1. As an example, we put $Q_{ab}$'s for $N=4$ in a matrix:
\begin{equation}
Q_{ab}=\left( \begin{array}{cccc} 
Q_{1}&Q_{1}\, Q_{\tau_{2}}Q_{\tau_{3}}Q_{\tau_{4}}&Q_{1}\,Q_{\tau_{3}}Q_{\tau_{4}}&Q_{1}\,Q_{\tau_{4}}\\
Q_{2}\,Q_{\tau_{1}}&Q_{2}&Q_{2}\,Q_{\tau_{1}}Q_{\tau_{3}}Q_{\tau_{4}}&Q_{2}\,Q_{\tau_{1}}Q_{\tau_{4}}\\
Q_{3}\,Q_{\tau_{1}}Q_{\tau_{2}}&Q_{3}\,Q_{\tau_{2}}&Q_{3}&Q_{3}\,Q_{\tau_{1}}Q_{\tau_{2}}Q_{\tau_{4}}\\
Q_{4}\,Q_{\tau_{1}}Q_{\tau_{2}}Q_{\tau_{3}}&Q_{4}\,Q_{\tau_{2}}Q_{\tau_{3}}&Q_{4}\,Q_{\tau_{3}}&Q_{4}
\end{array}\right).\end{equation}
\begin{figure}[h]
  \begin{center}
    \includegraphics[width=0.4\textwidth]{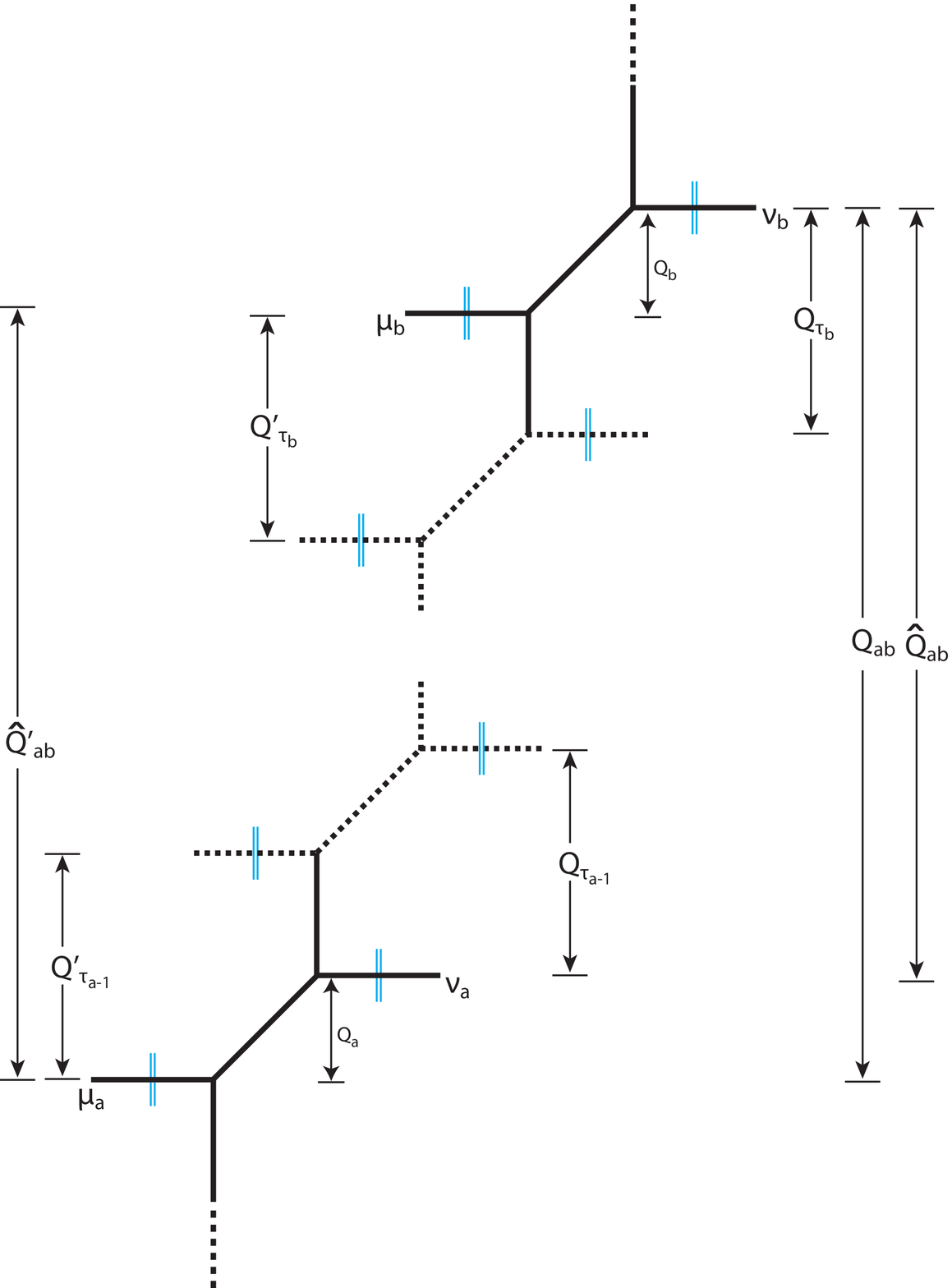}
  \end{center}
  \caption{The pictorial representation of the K\"{a}hler parameters used in the partition function.}\label{twostrip2}
\end{figure}
Using Eq.(\ref{id}), the numerator can be written only in terms of $Q_{ab}$
\begin{align}
\prod_{a,b=1}^{N}\prod_{i,j,k=1}^{\infty}(1-Q_{\tau}^{k-1}Q_{ab}\,t^{\mu_{a,i}-j+1/2}q^{\nu_{b,j}^{t}-i+1/2})(1-Q_{\tau}^{k}Q_{ab}^{-1}\,t^{\nu_{b,i}-j+1/2}q^{\mu_{a,j}^{t}-i+1/2}).
\end{align}
The denominator of the partition function for the $A_{N-1}$ singularity has the form
\begin{align}
\prod_{a,b=1}^{N}\prod_{i,j,k=1}^{\infty}(1-Q_{\tau}^{k-1}\widetilde{Q}'_{ab}\,t^{\mu_{a,i}-j}q^{\mu_{b,j}^{t}-i+1})^{-1}(1-Q_{\tau}^{k-1}\widetilde{Q}_{ab}\,t^{\nu_{a,i}-j+1}q^{\nu_{b,j}^{t}-i})^{-1},
\end{align}
where the K\"{a}hler parameters $\widetilde{Q}_{ab}$ are defined as follows:
\begin{equation}\label{eq:Qabtilde}
\widetilde{Q}_{ab}=\left\{ \begin{array}{lll}\prod_{i=b}^{a-1}Q_{\tau_{i}},& \mbox{for}\, a> b,\\ 
					Q_{\tau},& \mbox{for}\, a=b,\\ 
                                            Q_{\tau}/\prod_{i=a}^{b-1}Q_{\tau_{i}},&\mbox{for}\, a<b, \end{array}\right.
\end{equation}
and $\widetilde{Q}'_{ab}$'s are defined by replacing $Q_{\tau_{i}}$'s by $Q'_{\tau_{i}}$. There is a simple relation between $\widetilde Q_{ab}$ and $\widetilde Q'_{ab}$: 
\begin{align}
\widetilde{Q}'_{ab}=\frac{Q_{a}}{Q_{b}}\widetilde{Q}_{ab}.
\end{align}
We prefer to use this somewhat redundant notation, since it will prove convenient later in our discussion. The constraints we need to impose on the K\"{a}hler parameters of different strips will be more transparent in this convention. Combining the contributions from numerator and denominator, we see that the partition function for M$_A $-strings is constructed out of the following infinite products: 
\begin{align}\label{eq:strippartition}
\prod_{a,b=1}^{N}\prod_{i,j,k=1}^{\infty}\frac{(1-Q_{\tau}^{k-1}Q_{ab}\,t^{\mu_{a,i}-j+1/2}q^{\nu_{b,j}^{t}-i+1/2})(1-Q_{\tau}^{k}Q_{ab}^{-1}\,t^{\nu_{b,i}-j+1/2}q^{\mu_{a,j}^{t}-i+1/2})}{(1-Q_{\tau}^{k-1}\widetilde{Q}'_{ab}\,t^{\mu_{a,i}-j}q^{\mu_{b,j}^{t}-i+1})(1-Q_{\tau}^{k-1}\widetilde{Q}_{ab}\,t^{\nu_{a,i}-j+1}q^{\nu_{b,j}^{t}-i})}.
\end{align}
The partition function of the $U(1)$ partition function that lives on a single M5 brane can be computed from the above expression by setting all the Young diagrams to be trivial, in other words, considering the closed amplitude,
\begin{align}
Z_{U(1)}=\prod_{a,b=1}^{N}\prod_{i,j,k=1}^{\infty}\frac{(1-Q_{\tau}^{k-1}Q_{ab}\,t^{-j+1/2}q^{-i+1/2})(1-Q_{\tau}^{k}Q_{ab}^{-1}\,t^{-j+1/2}q^{-i+1/2})}{(1-Q_{\tau}^{k-1}\widetilde{Q}'_{ab}\,t^{-j}q^{-i+1})(1-Q_{\tau}^{k-1}\widetilde{Q}_{ab}\,t^{-j+1}q^{-i})},
\end{align}
where the all the pre-factors are trivial.

\subsection{Domain wall partition function}
\label{sec:domainwall}

\par{We are interested in the partition functions of M-strings on $A_{N-1}$ singularity. Therefore, the contributions from single M5-branes, $Z_{U(1)}$'s, need to be factored out.  We normalize the open topological partition function by the closed one using the following identity: }
\begin{align}
\prod_{i,j=1}^{\infty}\frac{1-Q\,q^{\nu_{i}-j}t^{\mu_{j}^{t}-i+1}}{1-Q\,q^{-j}t^{-i+1}}=\prod_{(i,j)\in\nu}\Big(1-Q\,q^{\nu_{i}-j}t^{\mu_{j}^{t}-i+1}\Big)\prod_{(i,j)\in\mu}\Big(1-Q\,q^{-\mu_{i}+j-1}t^{-\nu_{j}^{t}+i}\Big).
\end{align}
\par{As in  \cite{Haghighat:2013gba}, we can define the partition function of a domain wall from the normalized topological string partition function:}
\begin{align}\nonumber
D_{\nu_{1}\mathellipsis\nu_{N}}^{\mu_{1}\mathellipsis\mu_{N}}&(Q_{\tau},Q_{ab},\widetilde{Q}_{ab},\widetilde{Q}'_{ab};\epsilon_{1},\epsilon_{2})\equiv\prod_{a=1}^{N}q^{-\frac{\Arrowvert\mu_{a}^{t}\Arrowvert^{2}}{2}}\widetilde{Z}_{\mu_{a}^{t}}(t^{-1},q^{-1})t^{-\frac{\Arrowvert\nu_{a}\Arrowvert^{2}}{2}}\widetilde{Z}_{\nu_{a}}(q^{-1},t^{-1})\\\nonumber
\times\prod_{a,b=1}^{N}\prod_{k=1}^{\infty}&\prod_{(i,j)\in\mu_{a}}\frac{(1-Q_{\tau}^{k-1}Q_{ab}\,t^{\mu_{a,i}-j+1/2}q^{\nu_{b,j}^{t}-i+1/2})(1-Q_{\tau}^{k}Q_{ab}^{-1}\,t^{-\mu_{a,i}+j-1/2}q^{-\nu_{b,j}^{t}+i-1/2})}{(1-Q_{\tau}^{k-1}\widetilde{Q}'_{ab}\,t^{\mu_{a,i}-j}q^{\mu_{b,j}^{t}-i+1})(1-Q_{\tau}^{k-1}\widetilde{Q}'_{ba}t^{-\mu_{a,i}+j-1}q^{-\mu_{b,j}^{t}+i})}\\
\times&\prod_{(i,j)\in\nu_{b}}\frac{(1-Q_{\tau}^{k-1}Q_{ab}\,t^{-\nu_{b,i}+j-1/2}q^{-\mu_{a,j}^{t}+i-1/2})(1-Q_{\tau}^{k}Q_{ab}^{-1}\,t^{\nu_{b,i}-j+1/2}q^{\mu_{a,j}^{t}-i+1/2})}{(1-Q_{\tau}^{k-1}\widetilde{Q}_{ba}\,t^{\nu_{b,i}-j+1}q^{\nu_{a,j}^{t}-i})(1-Q_{\tau}^{k-1}\widetilde{Q}_{ab}\,t^{-\nu_{b,i}+j}q^{-\nu_{a,j}^{t}+i-1})}.\label{eq:ANDomain}
\end{align}
Note that in this expression we have restored the factors of $ \widetilde Z_\mu(t^{-1},q^{-1}) $ that were left out of the previous discussion. The domain wall defined in \cite{Haghighat:2013gba} is just the special case of the new domain wall for $N=1$. For general $N$, we give the following interpretation for the quantity defined in \eqref{eq:ANDomain}: 
The ground states of the theory of $k$ M2 branes on flat transverse space on $T^2$ taking the size of $T^2$ to be much smaller than the length between the M5-branes (when the scalars are massed up) are labeled by a single Young
diagram of size $k$  \cite{Berman:2009xd,Lin:2004nb,Gomis:2008vc,Kim:2010mr}.    Here we have a situation where the transverse space to the M2 branes
has an $A_{N-1}$ singularity.  To describe the M2 branes in this geometry we need to place
$N$ copies of them before orbifolding the flat transverse space.  In particular this implies that
the configuration of ground states of M2 branes is characterized by $N$ Young diagrams with
total number of boxes being $k$. 
From this viewpoint the low energy modes of this system are given by a quantum mechanical system, where the Hilbert space  is formed by an $N$-tuple of Young diagrams
\begin{equation}
	\vec{\mu} = (\mu_1,\cdots,\mu_N),
\end{equation}
with the identity operator $I = \sum_{\mu} |\vec{\mu}\rangle\langle \vec{\mu}^t|$ and Hamiltonian $H = |\vec{\mu}|$. The Hamiltonian can again be interpreted as M2-brane mass where the size of $T^2$ times the tension of the M2 brane have been normalized to $1$ and $|\vec{\mu}|$ is their number. 
The domain wall arises by having the M2-branes ending
on M5-branes on either side of it.  Thus we can view the M5-brane as an operator acting
from the left vacua, labeled by $N$ partitions to right vacua, again labeled by $N$ partitions.
In other words, equation \eqref{eq:ANDomain} gives the matrix elements of this operator for this quantum mechanical system.

\subsection{Partition function of M5-branes on transverse $ A_{N-1} $ singularity}
\label{sec:mapartition}
In this section we assemble the contributions from the different strips that compose the toric geometry derived from Figure \ref{fig:braneweb2}, and we arrive at an expression for the refined topological string partition function corresponding to it (equation \eqref{eq:partitionfunction}). As discussed in section \ref{sec:s1s1}, this is also the partition function of the system of $ M $ M5-branes on transverse $ A_{N-1} $ singularity. More precisely, the partition function we compute is normalized by the contributions of the BPS states that do not arise from M2-branes stretching between the M5-branes (although these factors can be easily restored); our final expression is organized as a sum of contributions from different numbers of $ \mathrm{M}_{\mathrm{A}} $-strings wrapping the torus in the worldvolume of the M5-branes.

\par
\begin{figure}[h]
  \begin{center}
    \includegraphics[width=0.7\textwidth]{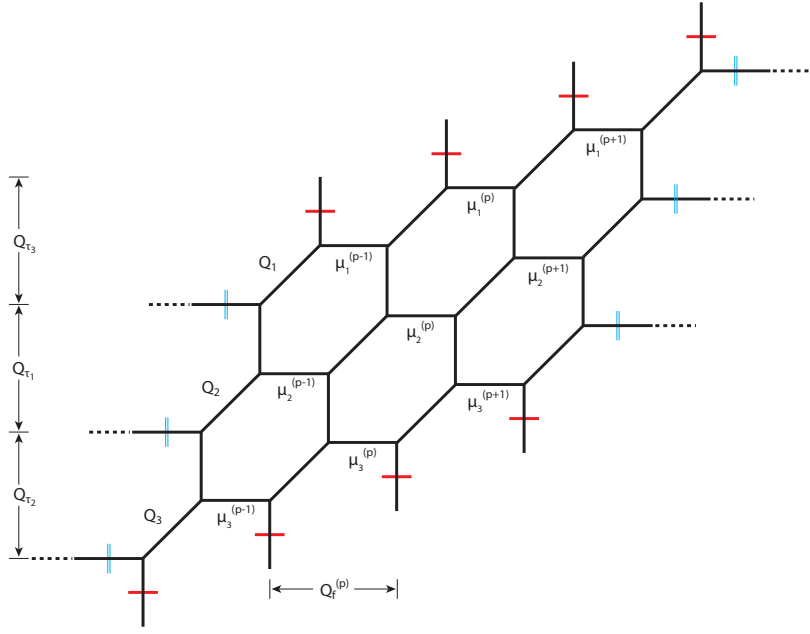}
  \end{center}
  \caption{Several domain walls are glued together. The small single (red) lines indicate that one has to glue along the vertical direction. This geometry depicts the $A_{2}$ singularity.}\label{glued}
\end{figure}

{In \cite{Haghighat:2013gba}, the normalized topological string partition function is recast in terms of the normalized Jacobi $\theta$-function
\begin{align}
\theta_{1}(\tau;z)=-ie^{i\pi\,\tau/4}e^{i\pi\,z}\prod_{k=1}^{\infty}(1-e^{2\pi i\,k\,\tau})(1-e^{2\pi i\,k\,\tau}e^{2\pi i\,z})(1-e^{2\pi i\,(k-1)\,\tau}e^{-2\pi i\,z}).
\end{align}

We will show that in the present, more general setup, the partition function can still be expressed in terms of $\theta$-functions. We first need to glue the building blocks together, Figure \ref{glued}. }

\par{The topological string partition functions in the presence of $M$ parallel M5-branes can be computed using the domain walls:
\begin{align}\nonumber
Z_M^{A_{N-1}}&=\sum_{\{\vec{\mu}^{(p)}\}_{p=1}^{M-1}} \left(\prod_{s=1}^{M-1}\prod_{a=1}^{N}\left(-Q_{f,a}^{(s)}\right)^{|\mu^{(s)}_{a}|}\right )\\ \nonumber
&\times D_{\mu^{(1)}_{1}\mathellipsis\mu_{N}^{(1)}}^{\emptyset\mathellipsis\emptyset}(Q_{\tau},Q^{(1)}_{ab},\widetilde{Q}^{(1)}_{ab},\widetilde{Q}'^{(1)}_{ab};\epsilon_{1},\epsilon_{2})D^{\mu^{(1)}_{1}\mathellipsis\mu_{N}^{(1)}}_{\mu^{(2)}_{1}\mathellipsis\mu_{N}^{(2)}}(Q_{\tau},Q_{ab}^{(2)},\widetilde{Q}^{(2)}_{ab},\widetilde{Q}'^{(2)}_{ab};\epsilon_{1},\epsilon_{2})\times\mathellipsis \\ &\times 
 D^{\mu^{(M-1)}_{1}\mathellipsis\mu_{N}^{(M-1)}}_{\emptyset\mathellipsis\emptyset}(Q_{\tau},Q_{ab}^{(M)},\widetilde{Q}_{ab}^{(M)},\widetilde{Q}'^{(M)}_{ab};\epsilon_{1},\epsilon_{2}),\label{eq:ZDomains}
\end{align}
where $\{\vec{\mu}^{(p)}\}$ denotes $N$-tuples of Young diagrams associated with gluing of the $ p^{th} $ and $ (p+1)^{st} $ domain walls. The K\"{a}hler parameters $Q^{(p)}_{ab},\widetilde{Q}^{(p)}_{ab}$ and $\widetilde{Q}'^{(p)}_{ab}$ belong to the $p^{th}$ domain wall whereas $Q_{f,a}^{(p)}$ denote the K\"{a}hler parameters along the fiber directions between $p^{th}$ and $(p+1)^{st}$ domain walls. Let us focus the contributions from the holomorphic maps which involve factors depending on $\vec{\mu}^{(p)}$. These are}

\begin{align}\nonumber
&\hspace{.32in}
\left.\begin{array}{r}
\prod_{a,b=1}^{N}\prod_{k=1}^{\infty}\prod_{(i,j)\in\mu^{(p)}_{b}}\frac{\left(1-Q_{\tau}^{k-1}Q^{(p)}_{ab}\,t^{-\mu^{(p)}_{b,i}+j-1/2}q^{-\mu_{a,j}^{(p-1),t}+i-1/2}\right)}{\left(1-Q_{\tau}^{k-1}\widetilde{Q}^{(p)}_{ba}\,t^{\mu^{(p)}_{b,i}-j+1}q^{\mu_{a,j}^{(p),t}-i}\right)}\\ \\
\times \frac{\left(1-Q_{\tau}^{k}\left(Q_{ab}^{(p)}\right)^{-1}\,t^{\mu_{b,i}^{(p)}-j+1/2}q^{\mu_{a,j}^{(p-1),t}-i+1/2}\right)}{\left(1-Q_{\tau}^{k-1}\widetilde{Q}^{(p)}_{ab}\,t^{-\mu^{(p)}_{b,i}+j}q^{-\mu_{a,j}^{(p),t}+i-1}\right)}
\end{array} \right \} \substack{\mbox{from the } \displaystyle{ p^{th}}\\\mbox{domain wall}}\\ \nonumber \\ 
&\hspace{.2in}
\left.\begin{array}{r}
\times\prod_{a,b=1}^{N}\prod_{k=1}^{\infty}\prod_{(i,j)\in\mu^{(p)}_{a}}\frac{\left(1-Q_{\tau}^{k-1}Q_{ab}^{(p+1)}\,t^{\mu^{(p)}_{a,i}-j+1/2}q^{\mu^{(p+1),t}_{b,j}-i+1/2}\right)}{\left(1-Q_{\tau}^{k-1}\widetilde{Q}'^{(p+1)}_{ab}\,t^{\mu^{(p)}_{a,i}-j}q^{\mu^{(p),t}_{b,j}-i+1}\right)}\\\\
\times\frac{\left(1-Q_{\tau}^{k}\left(Q_{ab}^{(p+1)}\right)^{-1}\,t^{-\mu^{(p)}_{a,i}+j-1/2}q^{-\mu^{(p+1),t}_{b,j}+i-1/2}\right)}{\left(1-Q_{\tau}^{k-1}\widetilde{Q}'^{(p+1)}_{ba}t^{-\mu^{(p)}_{a,i}+j-1}q^{-\mu^{(p),t}_{b,j}+i}\right)}
\end{array} \right \}\substack{\mbox{from the } \displaystyle{ (p+1)^{st}}\\\mbox{domain wall}}
\end{align}
and arise from the factors of
\[D_{\mu^{(p)}_{1}\mathellipsis\mu_{N}^{(p)}}^{\mu^{(p-1)}_{1}\mathellipsis\mu_{N}^{(p-1)}}\times D^{\mu^{(p)}_{1}\mathellipsis\mu_{N}^{(p)}}_{\mu^{(p+1)}_{1}\mathellipsis\mu_{N}^{(p+1)}} \]
appearing in \eqref{eq:ZDomains}. 

To compute the topological string partition function of the toric geometry we are required to glue together the $ M $ vertical strips along the horizontal edges. This imposes the constraint
\begin{align}
\widetilde{Q}'^{(p+1)}_{ab}=\widetilde{Q}^{(p)}_{ab}
\end{align}
which is easy to see from the toric diagrams. This is equivalent to imposing
\begin{align}
Q'^{(p+1)}_{\tau_{a}}=Q^{(p)}_{\tau_{a}}.
\end{align}
After imposing the gluing restrictions, using the fact that $a$ and $b$ are dummy variables, we see that the $ \vec{\mu}^{(p)} $-dependent terms can be written as
\larger[-2]
\begin{align}\nonumber\\ \nonumber
\prod_{a,b=1}^{N}\prod_{k=1}^{\infty}\prod_{(i,j)\in\mu^{(p)}_{a}}&\frac{\big(1-Q_{\tau}^{k-1}Q^{(p)}_{ba}\,t^{-\mu^{(p)}_{a,i}+j-1/2}q^{-\mu_{b,j}^{(p-1),t}+i-1/2}\big)\big(1-Q_{\tau}^{k}\left(Q_{ba}^{(p)}\right)^{-1}\,t^{\mu_{a,i}^{(p)}-j+1/2}q^{\mu_{b,j}^{(p-1),t}-i+1/2}\big)}{\big(1-Q_{\tau}^{k-1}\widetilde{Q}^{(p)}_{ab}\,t^{\mu^{(p)}_{a,i}-j+1}q^{\mu_{b,j}^{(p),t}-i}\big)\big(1-Q_{\tau}^{k-1}\widetilde{Q}^{(p)}_{ba}\,t^{-\mu^{(p)}_{a,i}+j}q^{-\mu_{b,j}^{(p),t}+i-1}\big)}\\
\times&\frac{\big(1-Q_{\tau}^{k-1}Q_{ab}^{(p+1)}\,t^{\mu^{(p)}_{a,i}-j+1/2}q^{\mu^{(p+1),t}_{b,j}-i+1/2}\big)\big(1-Q_{\tau}^{k}\left(Q_{ab}^{(p+1)}\right)^{-1}\,t^{-\mu^{(p)}_{a,i}+j-1/2}q^{-\mu^{(p+1),t}_{b,j}+i-1/2}\big)}{\big(1-Q_{\tau}^{k-1}\widetilde{Q}^{(p)}_{ab}\,t^{\mu^{(p)}_{a,i}-j}q^{\mu^{(p),t}_{b,j}-i+1}\big)\big(1-Q_{\tau}^{k-1}\widetilde{Q}^{(p)}_{ba}t^{-\mu^{(p)}_{a,i}+j-1}q^{-\mu^{(p),t}_{b,j}+i}\big)}.
\label{eq:pTerms}\end{align}
\normalsize
It is clear that numerator can be rewritten in terms of two $\theta$-functions:
\begin{align}
\mbox{Numerator}= A^{(p)} \cdot \prod_{a,b=1}^{N}\prod_{(i,j)\in\mu_{a}^{(p)}}\theta_{1}(\tau;z^{(p)}_{ab}(i,j))\theta_{1}(\tau;w^{(p)}_{ab}(i,j)),\label{eq:numerator}
\end{align}
where we have defined the arguments of the $\theta$-functions $z_{ab}^{(p)}(i,j)$ and $w_{ab}^{(p)}(i,j)$ as
\begin{align}
e^{2\pi i\,z_{ab}^{(p)}(i,j)}&\equiv \left(Q_{ab}^{(p+1)}\right)^{-1}\,t^{-\mu^{(p)}_{a,i}+j-1/2}q^{-\mu^{(p+1),t}_{b,j}+i-1/2},\\
e^{2\pi i\,w_{ab}^{(p)}(i,j)}&\equiv \left(Q_{ba}^{(p)}\right)^{-1}\,t^{\mu_{a,i}^{(p)}-j+1/2}q^{\mu_{b,j}^{(p-1),t}-i+1/2},
\end{align}
and
\begin{align} A^{(p)} = \prod_{a=1}^N \left[-e^{\pi i \tau/2}\prod_{k=1}^\infty (1-e^{2\pi ik\tau})^2\right]^{-N\vert \mu_a^{(p)}\vert}\prod_{b=1}^N\prod_{(i,j)\in\mu_a^{(p)}}e^{-\pi i (z_{ab}^{(p)}(i,j)+w_{ab}^{(p)}(i,j))}.\end{align}

\par{Let us now turn to the factors in the denominator of equation \eqref{eq:pTerms}. First of all, the factors for which $ a=b $ combine with the prefactors 
\begin{align}\label{tilde}\nonumber
&q^{-\frac{||\mu_a^{(p),t}||^2}{2}}t^{-\frac{||\mu_a^{(p)}||^2}{2}}\widetilde{Z}_{\mu_{a}^{(p),t}}(t^{-1},q^{-1})\widetilde{Z}_{\mu_{a}^{(p)}}(q^{-1},t^{-1})\\
&=(-1)^{|\mu_a^{(p)}|}\left(\frac{t}{q}\right)^{\frac{|\mu_a^{(p)}|}{2}}\prod_{(i,j)\in\mu^{(p)}_{a}}\frac{1}{1-q^{\mu_{a,j}^{(p),t}-i}t^{\mu^{(p)}_{a,i}-j+1}}\frac{1}{1-t^{-\mu_{a,i}^{(p)}+j}q^{-\mu_{a,j}^{(p),t}+i-1}}
\end{align}
to give
\begin{align}\nonumber
\left(-\sqrt{\frac{t}{q}}\right)^{\sum_a|\mu_a^{(p)}|}\prod_{a=1}^{N}\prod_{k=1}^{\infty}\prod_{(i,j)\in\mu^{(p)}_{a}}&\frac{1}{(1-Q_{\tau}^{k}\,t^{\mu^{(p)}_{a,i}-j}q^{\mu^{(p),t}_{a,j}-i+1})(1-Q_{\tau}^{k}\,t^{-\mu^{(p)}_{a,i}+j-1}q^{-\mu^{(p),t}_{a,j}+i})}\\
&\hspace{-.5in}\times\frac{1}{(1-Q_{\tau}^{k-1}\,t^{\mu^{(p)}_{a,i}-j+1}q^{\mu_{a,j}^{(p),t}-i})(1-Q_{\tau}^{k-1}\,t^{-\mu^{(p)}_{a,i}+j}q^{-\mu_{a,j}^{(p),t}+i-1})}.\label{eq:aDenom}
\end{align}
When $a\neq b$, we will need the following identity which follows from the definition of $\widetilde{Q}_{ab}$:
\begin{align}
\widetilde{Q}^{(p)}_{ab}\widetilde{Q}^{(p)}_{ba}=Q_{\tau}.
\end{align}
This allows us to write the denominator terms for $ a\neq b $ as
\begin{align}\nonumber
\prod_{a\neq b}^{N}\prod_{k=1}^{\infty}\prod_{(i,j)\in\mu^{(p)}_{a}}&\frac{1}{\big(1-Q_{\tau}^{k}\,\left(\widetilde{Q}^{(p)}_{ba}\right)^{-1}\,t^{\mu^{(p)}_{a,i}-j}q^{\mu^{(p),t}_{b,j}-i+1}\big)\big(1-Q_{\tau}^{k}\,\left(\widetilde{Q}_{ab}^{(p)}\right)^{-1}\,t^{-\mu^{(p)}_{a,i}+j-1}q^{-\mu^{(p),t}_{b,j}+i}\big)}\\
\times&\frac{1}{(1-Q_{\tau}^{k-1}\widetilde{Q}^{(p)}_{ab}\,t^{\mu^{(p)}_{a,i}-j+1}q^{\mu_{b,j}^{(p),t}-i})(1-Q_{\tau}^{k-1}\widetilde{Q}^{(p)}_{ba}\,t^{-\mu^{(p)}_{a,i}+j}q^{-\mu_{b,j}^{(p),t}+i-1})}.\label{eq:abDenom}
\end{align}
If we now define a new variable $\widehat{Q}^{(p)}_{ab}$ by
\begin{equation}\label{eq:Qabhat}
\widehat{Q}^{(p)}_{ab}=\left\{\begin{array}{ll}1,\qquad&\mbox{for}\,a=b\\ \widetilde{Q}^{(p)}_{ab}, &\mbox{for}\,a\neq b\end{array},\right.
\end{equation}
we can write the product of equations \eqref{eq:aDenom} and \eqref{eq:abDenom} as
\begin{align}
B^{(p)} \cdot \prod_{a,b=1}^N\prod_{(i,j)\in \mu_a^{(p)}}\frac{1}{\theta_{1}(\tau;u_{ab}^{(s)}(i,j))\theta_{1}(\tau;v_{ab}^{(s)}(i,j))},
\end{align}
where
\begin{align}
e^{2\pi i\,u_{ab}^{(p)}(i,j)}&\equiv \left(\widehat{Q}^{(p)}_{ba}\right)^{-1}\, t^{\mu^{(p)}_{a,i}-j}q^{\mu^{(p),t}_{b,j}-i+1}\\
e^{2\pi i\,v_{ab}^{(p)}(i,j)}&\equiv \left(\widehat{Q}_{ab}^{(p)}\right)^{-1}\,t^{-\mu^{(p)}_{a,i}+j-1}q^{-\mu^{(p),t}_{b,j}+i}
\end{align}
and
\[ B^{(p)} = \prod_{a=1}^N \left(-\sqrt{\frac{t}{q}}\right)^{|\mu_a^{(p)}|}\left[-e^{\pi i \tau/2}\prod_{k=1}^\infty (1-e^{2\pi ik\tau})^2\right]^{N\vert \mu_a^{(p)}\vert}\prod_{b=1}^N\prod_{(i,j)\in\mu^{(p)}_{a}}e^{\pi i\,u_{ab}^{(p)}(i,j)+\pi i\,v_{ab}^{(p)}(i,j)}.\]
Therefore equation \eqref{eq:ZDomains} simplifies to the following expression:
\begin{align}\nonumber
Z_M^{A_{N-1}}&=\sum_{\{\vec{\mu}^{(p)}\}_{p=1}^{M-1}} \prod_{s=1}^{M-1} C^{(s)} \prod_{a=1}^{N}\left(-Q_{f,a}^{(s)}\right)^{|\mu^{(s)}_{a}|} \prod_{(i,j)\in\mu_{a}^{(s)}}\prod_{b=1}^{N}\frac{\theta_{1}(\tau;z_{ab}^{(s)}(i,j))\theta_{1}(\tau;w_{ab}^{(s)}(i,j))}{\theta_{1}(\tau;u_{ab}^{(s)}(i,j))\theta_{1}(\tau;v_{ab}^{(s)}(i,j))}.
\end{align} 
It remains to simplify the prefactor
\begin{align} C^{(p)} =\prod_{a=1}^N\left(-\sqrt{\frac{t}{q}}\right)^{|\mu_a^{(p)}|}\prod_{b=1}^{N}\prod_{(i,j)\in\mu^{(p)}_{a}}e^{-\pi i\,z_{ab}^{(p)}(i,j)-w_{ab}^{(p)}(i,j)}e^{\pi i\,u_{ab}^{(p)}(i,j)+v_{ab}^{(p)}(i,j)}.\end{align}
First of all, we have
\begin{align}
&\prod_{a,b=1}^{N}\prod_{(i,j)\in\mu^{(p)}_{a}}e^{-\pi i\,z_{ab}^{(p)}(i,j)-w_{ab}^{(p)}(i,j)}=\left(\prod_{a,b=1}^{N}\prod_{(i,j)\in\mu^{(p)}_{a}}Q^{(p+1)}_{ab}Q^{(p)}_{ba}\,q^{\mu^{(p+1),t}_{b,j}-\mu_{b,j}^{(p-1),t}}\right)^{1/2}.
\end{align}
We can simplify this expression by noting that
\begin{equation}
Q_{ab}^{(p+1)}Q_{ba}^{(p)}=\left\{\begin{array}{ll}
Q_{b}^{(p)}Q_{b}^{(p+1)},\qquad &\mbox{for}\,a=b,\\
Q_{\tau}Q_{b}^{(p)}Q_{b}^{(p+1)},&\mbox{for}\,a\neq b,
\end{array}\right.
\end{equation}
so that
\begin{align}\label{factors}
\bigg(\prod_{a,b=1}^{N}\prod_{(i,j)\in\mu^{(p)}_{a}}Q^{(p+1)}_{ab}Q^{(p)}_{ba}\bigg)^{1/2}=\prod_{a=1}^NQ_{\tau}^{\frac{N-1}{2}|\mu_{a}^{(p)}|}\prod_{b=1}^{N}\left(Q_{b}^{(p+1)}Q_{b}^{(p)} \right)^{\frac{1}{2}|\mu_{a}^{(p)}|}.
\end{align}
Furthermore, it turns out that the $ q- $dependent terms in equation \eqref{factors} all cancel. To see this, let us isolate the factors associated to the $ p^{th} $ and $(p+1)^{st}$ four-cycles:
\begin{align}
\prod_{(i,j)\in\mu^{(p)}_{a}}q^{\mu^{(p+1),t}_{b,j}-\mu_{b,j}^{(p-1),t}}\prod_{(i,j)\in\mu^{(p+1)}_{a}}q^{\mu^{(p+2),t}_{b,j}-\mu_{b,j}^{(p),t}}.
\end{align}
Using the identity $ \sum_{(i,j)\in\nu}\mu_{j}^{t}=\sum_{(i,j)\in\mu}\nu_{j}^{t}$,
this simplifies to
\begin{align}
\prod_{(i,j)\in\mu^{(p)}_{a}}q^{-\mu_{b,j}^{(p-1),t}}\prod_{(i,j)\in\mu^{(p+1)}_{a}}q^{\mu^{(p+2),t}_{b,j}}.
\end{align}
By applying this identity at each four-cycle, we can cancel all $q$-dependent factors in equation \eqref{factors} against each other.

Likewise, one can show that
\begin{align}
\prod_{a,b=1}^{N}\prod_{(i,j)\in\mu^{(p)}_{a}}e^{\pi i\,u_{ab}^{(p)}(i,j)+\pi i\,v_{ab}^{(p)}(i,j)}=\prod_{a=1}^N\left(\frac{t}{q}\right)^{-N\frac{|\mu_a^{(p)}|}{2}}Q_{\tau}^{-\frac{N-1}{2}|\mu_{a}^{(p)}|},
\end{align}
and therefore
\[ C^{(p)} = \prod_{a=1}^N\left(-\left(\frac{q}{t}\right)^{\frac{(N-1)}{2}}\prod_{b=1}^{N}\left(Q_{b}^{(p+1)}Q_{b}^{(p)}\right)^{1/2}
\right)^{|\mu_a^{(p)}|}.\]
Finally, we define
\begin{align} \overline{Q}^{(p)}_{f,a}:= e^{2\pi i\overline{t}_{f,a}^{(p)}} \equiv \left(\frac{q}{t}\right)^{\frac{(N-1)}{2}}Q^{(p)}_{f,a} \prod_{b=1}^{N}\left(Q_{b}^{(p+1)}Q_{b}^{(p)}\right)^{1/2} =  \left(\frac{q}{t}\right)^{\frac{(N-1)}{2}}Q^{(p)}_{f,a} Q_m^N, \end{align}
where we have set
\begin{equation}
	Q_m = e^{2\pi i m} = \left(\prod_{b=1}^n Q_b^{(p)}\right)^{\frac{1}{N}}.
\end{equation}
Here we note that $m$ corresponds to the physical mass parameter introduced in section \ref{sec:basics}, and that its definition is in fact independent of the label $p$.
We obtain a very compact final expression for the partition function of $ M $ M5-branes on transverse $ A_{N-1} $ singularity:
\begin{align} \label{eq:partitionfunction}
\boxed{Z_M^{A_{N-1}} = \sum_{\{\vec{\mu}^{(p)} \}_{p=1}^{M-1}}\prod_{s=1}^{M-1}\prod_{a=1}^{N}\left(\bar{Q}^{(s)}_{f,a}\right)^{|\mu_{a}^{(s)}|}\prod_{(i,j)\in\mu_{a}^{(s)}}\prod_{b=1}^{N}\frac{\theta_{1}(\tau;z_{ab}^{(s)}(i,j))\theta_{1}(\tau;w_{ab}^{(s)}(i,j))}{\theta_{1}(\tau;u_{ab}^{(s)}(i,j))\theta_{1}(\tau;v_{ab}^{(s)}(i,j))}.}
\end{align}
Remember that the partition function $Z_M^{A_{N-1}}$ is normalized by the partition functions of single M5 branes, $Z_{U(1)}^{(p)}$. The K\"{a}hler parameters of each domain wall are different, therefore the overall normalization is by $\prod_{p=1}^{M}Z_{U(1)}^{(p)}$.
For convenience, let us collect here the following definitions which were given in the previous discussion:
\begin{align}
e^{2\pi i\,z_{ab}^{(p)}(i,j)}&\equiv \left( Q_{ab}^{(p+1)}\right)^{-1}\,t^{-\mu^{(p)}_{a,i}+j-1/2}q^{-\mu^{(p+1),t}_{b,j}+i-1/2}\nonumber\\
e^{2\pi i\,w_{ab}^{(p)}(i,j)}&\equiv \left( Q_{ba}^{(p)}\right)^{-1}\,t^{\mu_{a,i}^{(p)}-j+1/2}q^{\mu_{b,j}^{(p-1),t}-i+1/2}\nonumber\\
e^{2\pi i\,u_{ab}^{(p)}(i,j)}&\equiv \left(\widehat{Q}^{(p)}_{ba}\right)^{-1}\, t^{\mu^{(p)}_{a,i}-j}q^{\mu^{(p),t}_{b,j}-i+1}\nonumber\\
e^{2\pi i\,v_{ab}^{(p)}(i,j)}&\equiv \left(\widehat{Q}_{ab}^{(s)}\right)^{-1}\,t^{-\mu^{(p)}_{a,i}+j-1}q^{-\mu^{(p),t}_{b,j}+i}.\nonumber
\end{align}
The K\"ahler parameters appearing here can all be expressed in terms of the parameters $Q_m = e^{2\pi i m}$, $Q^{(p)}_{\tau_a} = e^{2\pi i\tau^a_p}$, and $Q_{\tau}=e^{2\pi i \tau}$ which have the interpretation of mass-rotation, $SU(N)$ fugacities and elliptic parameter of the two-dimensional quiver theory discussed in section \ref{sec:quiver}. Finally, we define the parameters
\begin{equation}
	\bar{Q}_f^{(p)} = e^{2\pi i t_f^{(p)}} = \left(\prod_{a=1}^N \bar{Q}_{f,a}^{(p)}\right)^{\frac{1}{N}},
\end{equation}
which we identify with the tension of the M-strings, or equivalently the distances between the M5-branes. From the factor $\prod_{a=1}^N \left(\bar{Q}^{(p)}_{f,a}\right)^{|\mu_a^{(s)}|}$ in the partition function (\ref{eq:partitionfunction}) we can extract an overall factor
\begin{equation}
	\left(\bar{Q}_f^{(p)}\right)^{\sum_{a=1}^N |\mu_a^{(p)}|},
\end{equation}
which in the quantum mechanical framework introduced in section \ref{sec:domainwall} is associated to the propagator between the $p$-th and $(p+1)$-st domain wall. The remaining factor
will depend on the individual sizes of partitions $\mu_1^{(p)},\cdots,\mu_N^{(p)}$. These factors, which we henceforth denote by $R_{\vec{\mu}^{(p)}}$, should combine with the product over Jacobi $\theta$-functions in such a way that the partition function (\ref{eq:partitionfunction}) displays the expected modular properties. Let us discuss this in more detail. 

The Jacobi $\theta$-function acquires a non-trivial phase under the modular transformation
\begin{equation}
	\theta_1(-1/\tau;z/\tau)=  -i(-i\tau)^{1/2}\exp(\pi i z^2/\tau)\theta_1(\tau;z).
\end{equation}
This modular anomaly can be traced back to the appearance of the second Eisenstein series $E_2(\tau)$ in the following expression for $\theta_1(\tau,z)$:
\begin{equation}
	\theta_1(\tau,z) = \eta(\tau)^3 z \exp\left[\sum_{k\geq 1} \frac{B_{2k}}{(2k)(2k!)} E_{2k}(\tau) z^{2k}\right].
\end{equation}

As discussed in \cite{Haghighat:2013gba}, the modular anomaly can be traded for a holomorphic anomaly: 
this is achieved by replacing $E_2(\tau)$ by its modular completion $\widehat{E}_2(\tau,\bar{\tau}) = E_2(\tau) - \frac{3}{\pi \textrm{Im}(\tau)}$ in each occurrence of the $ \theta $-function, at the cost of introducing a mild dependence of the partition function on the anti-holomorphic parameter $ \bar \tau $. This leads to a modified partition function which has the following modular behaviour
\[ Z_M^{A_{N-1}}(\tau,\bar\tau; t_{f}^{(p)}; m,\tau^a_p,\epsilon_1,\epsilon_2)= Z_M^{A_{N-1}}(-1/\tau,-1/\bar\tau; t_{f}^{(p)}; m/\tau,\tau^a_p/\tau,\epsilon_1/\tau,\epsilon_2/\tau)\]
and satisfies a holomorphic anomaly equation. This equation relates derivatives of the partition function with respect to $\bar{\tau}$ to derivatives with respect to $t_f^{(p)}$. For this to be true it is critical that for each summand corresponding to a choice of partitions $\{\mu_a^{(p)}\}$ the coefficient of $\bar{\tau}$ is a function of the combinations $\sum_{a=1}^N |\mu_a^{(p)}|$ only. For this highly non-trivial statement to hold the residual factors $R_{\vec{\mu}^{(p)}}$ must combine with the product over the theta-functions appropriately.

\subsection{Direct computation of the $ \mathrm{M}_{\mathrm{A}} $-string elliptic genus}\label{sec:ell genus}
\label{sec:eg}
From formula (\ref{eq:partitionfunction}) we can extract the elliptic genus for $\textrm{M}_{\textrm{A}}$-strings arising from suspended M2-branes between $M$ M5-branes in the presence of $A_{N-1}$ singularity:
\begin{equation} \label{eq:eg}
	\textrm{Ell}(N,\vec{k})= \sum_{\sum_{a} |\mu_a^{(p)}|=k_p} \prod_{s=1}^{M-1} R_{\vec{\mu}^{(s)}} \prod_{a,b=1}^N \prod_{(i,j)\in \mu_a^{(s)} }\frac{\theta_1(\tau;z^{(s)}_{ab}(i,j))\theta_1(\tau;w_{ab}^{(s)}(i,j))}{\theta_1(\tau;u_{ab}^{(s)}(i,j))\theta_1(\tau;v_{ab}^{(s)}(i,j))},
\end{equation}
where $k_p, $ for $ p=1,\dots,M-1,$ is the number of M2-branes suspended between the $ p $-th and $ (p+1) $-st M5-brane. An alternative method to computing this elliptic genus would be through a detailed understanding of the bundles over the instanton moduli space \eqref{eq:instmoduli} of the 2d quiver gauge theory in which the fermions and bosons transform \cite{Haghighat:2013gba,Gritsenko}. In the present paper, we will content ourself with sketching this approach. The bosons are sections of the tangent bundle of $\mathcal{M}_{k_1,\cdots,k_{M-1}}$, whereas the fermions are sections of the bundle $V$ discussed in section \ref{sec:quiver}.
The weights of these bundles at the fixed points were worked out in \cite{CO}, and following \cite{Gritsenko} one can use them to compute the elliptic genus by employing the Hirzebruch-Riemann-Roch theorem as follows:
\begin{equation}
	\textrm{Ell}(N,\vec{k}) = \int_{\mathcal{M}_{k_1,\cdots,k_{M-1}}} \textrm{ch}(E_{Q_{\tau}}) \textrm{Td}(\mathcal{TM}_{k_1,\cdots,k_{M-1}}),
\end{equation}
where $\mathcal{TM}$ is the tangent bundle, and the bundle $E_{Q_{\tau}}$ is given by
\begin{equation} \label{eq:egintegral}
	E_{Q_{\tau}} = \bigotimes_{l=0}^{\infty} \bigwedge_{Q_{\tau}^{l-1}} V \bigotimes_{l=1}^{\infty} \bigwedge_{Q_{\tau}^l} V^* \otimes \bigotimes_{l=1}^{\infty} S_{Q_{\tau}^l} \mathcal{TM}^* \otimes \bigotimes_{l=1}^{\infty} S_{Q_{\tau}^l} \mathcal{TM},
\end{equation}
where for brevity we have suppressed the dependence of the bundle $E_{Q_{\tau}}$ on the different parameters. The fugacities on which the elliptic genus depends can be obtained from the quiver description of Section \ref{sec:quiver} as follows. For each node of the inner quiver we get $4N$ fugacities from the bifundamental fields $\tilde{Q}$, $Q$, $\Lambda^Q$ and $\Lambda^{\tilde{Q}}$, which are multiplied by $M-1$ as there are $M-1$ inner nodes. Furthermore, we have $3(M-1)$ parameters from the fields $\Lambda^{\Phi}$, $B$, $\tilde B$ associated to each inner node, and $4(M-2)$ fugacities from the bifundamentals $\Sigma, \Lambda^{B}$, $\Lambda^{\tilde B}$ and $\Lambda^{\Phi}$ of the inner quiver. Thus we have a total of $4N(M-1) + 3(M-1) + 4(M-2)$ parameters. However, there will be constraints from superpotentials and gauge anomalies. Including all these constraints should reduce the number of independent parameters to: 
\begin{equation}
	\#\textrm{fugacities} = NM - M + 3.
\end{equation}
Having directly computed the elliptic genus then makes it possible to reconstruct the partition function of M5-branes in the presence of $A_{N-1}$ singularity as follows: 
\begin{equation} \label{eq:M5partition}
	Z_{M5} = \left(\prod_{p=1}^{N}Z^{(p)}_{U(1)} \right) Z_M^{A_{N-1}} = \left(\prod_{p=1}^{N}Z^{(p)}_{U(1)} \right) \left(\sum_{\vec{k}} \prod_{s=1}^{M-1}  (\bar{Q}_{f}^{(s)})^{k_s} \textrm{Ell}(N, \vec{k})\right),
\end{equation}
where $Z_{U(1)}$ is the contribution of a single M5-brane to the partition function and does not contain any contributions from BPS string states.
Yet another way to compute the elliptic genus of  $\textrm{M}_\textrm{A}$-strings would be by directly applying the techniques developed in \cite{Benini:2013nda, Gadde:2013dda, Benini:2013xpa} to the 2d quiver gauge theory described in Section \ref{sec:quiver}. Here we will illustrate how this works in the case of $ k $ M2-branes suspended between two M5-branes on transverse $ \textrm{TN}_1 $ space, that is for $M=2, N=1$. This generalizes the result of \cite{Haghighat:2013gba} to arbitrary mass. On the one hand, the partition function for $ k $ M-strings is given by
\begin{align} Z^k_{M-strings}(\tau,m,\epsilon_1,\epsilon_2) = \sum_{\vert \nu \vert = k} \prod_{(i,j)\in \nu}\frac{\theta_1(\tau;Q_m^{-1}q^{\nu_i-j+1/2}t^{-i+1/2})\theta_1(\tau;Q_m^{-1}q^{-\nu_i+j-1/2}t^{i-1/2})}{\theta_1(\tau;q^{\nu_i-j+1}t^{\nu_j^t-i})\theta_1(\tau;q^{\nu_i-j}t^{\nu_j^t-i+1})}.\label{eq:kMstrings}\end{align}
On the other hand, the elliptic genus of the 2d affine $ A_1 $ quiver gauge theory with one node removed  and $ U(k) $ gauge group (see Figure \ref{fig:A1quiver})
\begin{figure}[here!]
  \centering
	\includegraphics[width=0.8\textwidth]{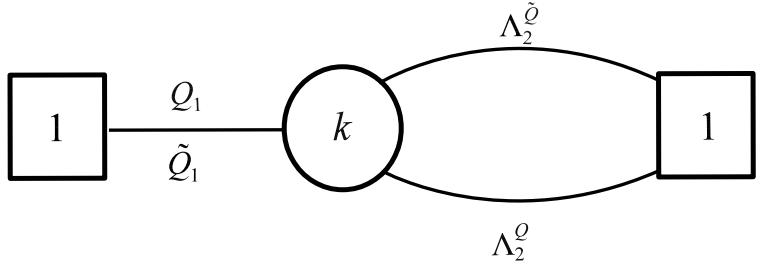}
  \caption{The quiver for two M5 branes in presence of transverse $\textrm{TN}_1$. }
  \label{fig:A1quiver}
\end{figure}
 is given by\footnote{We are grateful to A. Gadde for communicating this result to us.}
\begin{align}\nonumber \mbox{Ell}(k;\tau,m,\epsilon_1,\epsilon_2) &= \frac{1}{k!}\oint \frac{dz_\alpha}{2\pi iz_\alpha}\prod_{\alpha,\beta=1}^k\left(\frac{\theta_1(\tau;z_\alpha/z_\beta)\theta_1(\tau;Q_\tau tq^{-1} z_\alpha/z_\beta)}{\theta_1(\tau;q z_\alpha/z_\beta)\theta_1(\tau;t^{-1}z_\alpha/z_\beta)}\right)\\
&\hspace{0.88in} \times\prod_{\alpha=1}^k
\left(\frac{\theta_1(\tau;Q_mz_\alpha)\theta_1(\tau;Q_mz_\alpha^{-1})}{\theta_1(\tau;\sqrt{q/t}z_\alpha)\theta_1(\tau;\sqrt{q/t}z_\alpha)}\right)\nonumber\\
&= \sum_{\nu} \bigg(\prod_{\substack{(i_1,j_1)\in \nu\\(i_2,j_2)\in \nu}}\frac{\theta_1(\tau;q^{j_1-j_2+1}t^{-(i_1-i_2+1)})\theta_1(\tau;q^{j_1-j_2}t^{-(i_1-i_2)})}{\theta_1(\tau;q^{j_1-j_2}t^{-(i_1-i_2+1)})\theta_1(\tau;q^{j_1-j_2+1}t^{-(i_1-i_2)})}\nonumber\\
&\hspace{.35in}\times \prod_{(i,j)\in \nu}\frac{\theta_1(\tau;Q_m^{-1}q^{j-1/2}t^{-i+1/2})\theta_1(\tau;Q_m^{-1}q^{-j+1/2}t^{i-1/2})}{\theta_1(\tau;q^{j}t^{-i})\theta_1(\tau;q^{-j-1}t^{i-1})}\Bigg),
\label{eq:ellGenus}
\end{align}
where it is understood that each occurrence of $ \theta_1(\tau; 1) $ in the previous equation is to be replaced by $ -\partial_z\theta_1(\tau;z)\vert_{z=1}$.
The two expressions are superficially different, but one can show that for each Young diagram the product over pairs of boxes of Eqn. \eqref{eq:ellGenus} simplifies to the product over individual boxes of the same Young diagram in equation \eqref{eq:kMstrings}.  Analogously, we predict that the elliptic genus of 2d affine $ A_{M-1} $ quiver theories with $ N $ flavors will coincide with the partition function of M-strings for a system of $ M $ parallel M5-branes on transverse $ \textrm{TN}_{N} $ space.

\section{Concluding remarks}

In this paper we have shown that the partition function of $M$ parallel M5 branes in the presence
of transverse $A_{N-1}$ singularity compactified on $T^2$, can be computed for arbitrary
supersymmetry preserving twists using the corresponding strings, obtained by
stretched M2 branes suspended between M5 branes and wrapping $T^2$.  Moreover we have shown that their world volume
theory is given by 2d quiver gauge theory and that can be used to effectively compute
the partition function of this theory.  In a way, this is similar in spirit to quantum field theories
where the partition functions can be computed using the particle contributions to amplitudes.
Here the analog of particles are the strings and they indeed do yield the partition function
for the (1,0) superconformal theory at least when compactified on $T^2$.  Note that as a special
case of our computation we can also compute in this way the partition function of 6d $(2,0)$ A-type
theory.  Furthermore, since we can use this building block to compute the superconformal index
of the 6d theory \cite{Lockhart:2012vp,Kim:2012qf}, we have thus effectively related the superconformal index in 6d to the computation
of elliptic genera on the collection of 2d theories living on the resulting strings.  This reinforces
the picture that these 6d theories are indeed a theory of interacting strings.

In this paper we focused on A-type 6d (2,0) theory.  It is natural to ask what one can say
about the more general D or E-type (2,0) theory.  Even before putting this in the prsensence
of $A_{N-1}$ type singularity, the computation of their supersymmetric amplitudes are much
more difficult.  In principle one can do this using geometric engineering of the corresponding
theories, or by developing suitable instanton Calculus techniques to compute
the partition function of 5d lift of ${\cal N}=2^*$ theory for D or E type gauge theory.  On the other hand,
it is straight-forward to extend the analysis of this paper to these cases by placing
them in the presence of transverse $A_{N-1}$ singularity and obtain the associated
2d quiver theory.  The computation of partition functions of the associated
(2,0) and (1,0) superconformal theories in 6d reduces to the
computation of elliptic genera of the associated 2d quiver theories. We are currently pursuing this idea\footnote{Work in progress with A. Gadde.}.

\section*{Acknowledgements}
We would like to thank A. Gadde, D. Jafferis, N. Mekareeya, M. Ro\v{c}ek and R. Wimmer for useful discussions. We would like to thank the SCGP for hospitality during the 11th Simons Workshop on math
and physics, where this work was initiated.
 C.K. would also like to thank the Harvard University Theoretical High Energy Physics/String Theory group for hospitality.  

The work of B.H. is supported by NSF grant DMS-0804454. The work of C.V. is supported in part by NSF grant PHY-1067976.

\bibliography{references}

\end{document}